\documentclass[12pt,preprint]{emulateapj}
\usepackage{lscape}

\begin{document}

\title{Mapping the Shores of the Brown Dwarf Desert II: 
Multiple Star Formation in Taurus-Auriga}

\author{
Adam L. Kraus\altaffilmark{1}, 
Michael J. Ireland\altaffilmark{2}, 
Frantz Martinache\altaffilmark{3},
\& Lynne A. Hillenbrand\altaffilmark{4}
}

\altaffiltext{1}{Hubble Fellow; Institute for Astronomy, University of 
Hawaii, 2680 Woodlawn Dr., Honolulu, HI 96822, USA}
\altaffiltext{2}{Sydney Institute for Astronomy (SIfA), School of Physics, 
University of Sydney, NSW 2006, Australia}
\altaffiltext{3}{National Astronomical Observatory of Japan, Subaru 
Telescope, Hilo, HI 96720, USA}
\altaffiltext{4}{California Institute of Technology, Department of 
Astrophysics, MC 249-17, Pasadena, CA 91125, USA}

\begin{abstract}

We have conducted a high-resolution imaging study of the Taurus-Auriga 
star-forming region in order to characterize the primordial outcome of 
multiple star formation and the extent of the brown dwarf desert. Our 
survey identified 16 new binary companions to primary stars with masses of 
0.25-2.5 $M_{\sun}$, raising the total number of binary pairs (including 
components of high-order multiples) with separations of 3--5000 AU to 90. 
We find that $\sim$2/3--3/4 of all Taurus members are multiple systems of 
two or more stars, while the other $\sim$1/4--1/3 appear to have formed as 
single stars; the distribution of high-order multiplicity suggests that 
fragmentation into a wide binary has no impact on the subsequent 
probability that either component will fragment again. The separation 
distribution for solar-type stars (0.7--2.5 $M_{\sun}$) is nearly log-flat 
over separations of 3--5000 AU, but lower-mass stars (0.25--0.7 
$M_{\sun}$) show a paucity of binary companions with separations of 
$\ga$200 AU. Across this full mass range, companion masses are well 
described with a linear-flat function; all system mass ratios 
($q=M_B/M_A$) are equally probable, apparently including substellar 
companions. Our results are broadly consistent with the two expected modes 
of binary formation (freefall fragmentation on large scales and disk 
fragmentation on small scales), but the distributions provide some clues 
as to the epochs at which the companions are likely to form.

\end{abstract}


\section{Introduction}

The frequency and properties of multiple star systems offer powerful 
constraints on star formation and early cluster evolution. For example, 
the newest generation of theoretical models now broadly match the slope 
and turnover of the initial mass function (IMF; e.g., Bate 2009a). 
Simultaneous agreement with the mass-dependent frequency, separation 
distribution, and mass ratio distribution for binary systems is a far more 
demanding criterion, and one that has yet to be achieved. The ubiquity of 
binary systems suggests that an understanding of multiple star formation 
is also necessary to truly understand other processes like cluster 
formation, protoplanetary disk evolution, and planet formation.

The past two decades have seen numerous studies of nearby field binary 
systems in order to constrain their frequency and properties. These 
surveys (e.g. Duquennoy \& Mayor 1991, hereafter DM91; Fischer \& Marcy 
1992, hereafter FM92; Close et al. 2003; Bouy et al. 2003; Burgasser et 
al. 2003; Raghavan et al. 2010) have found that binary frequencies and 
properties are very strongly dependent on mass. Solar-mass stars have high 
binary frequencies ($\ga$60\%), the separation distribution appears to be 
log-normal with a peak of $\sim$30 AU and includes binary stars with 
separations of $>$10$^4$ AU, and the mass ratio distribution includes many 
low-mass companions. By contrast, very low-mass stars and brown dwarfs 
have a low binary frequency ($\sim$20\%), small mean separations ($\sim$4 
AU) and maximum separations ($\la$20 AU), and a strong tendency to have 
equal-mass companions. The form of the mass-dependent transition between 
these regimes is still unclear for field stars, though there is some 
evidence for a smooth transition (e.g. FM92; Reid \& Gizis 1997).

Parallel surveys of young star-forming regions have supported some of 
these conclusions, but also indicated intriguing differences. In surveys 
of Class II/III T Tauri stars in sparse associations like Taurus and Upper 
Sco, the companion frequency is very high ($\ga$80\%; Ghez et al. 1993; 
Simon et al. 1995; K\"ohler et al. 2000; Kraus et al. 2008). This trend 
does not seem to hold for denser young clusters like IC 348 and the ONC 
(Duch\^ene et al. 1999; K\"ohler et al. 2006; Reipurth et al. 2007) and 
old open clusters like $\alpha$ Per, the Pleiades, and Praesepe (Bouvier 
et al. 1997; Patience et al. 2002), where the binary frequency is similar 
to that of field stars; it is still unclear whether this difference is a 
primordial feature caused by different initial conditions or an 
evolutionary feature resulting from dynamical interactions. Sparse 
associations also have a much higher frequency of wide binary companions 
than either dense clusters or the field (Kraus \& Hillenbrand 2007a, 
2009a; Reipurth et al. 2007), a difference that most likely does result 
from dynamical disruption of wide binary systems in the latter 
populations. This strongly argues that neither clusters nor the field 
represent a dynamically pristine population, and therefore that they 
provide limited constraints on the binary formation process. By contrast, 
sparse associations seem to represent a more primordial population. 
Observations of younger Class 0/I T Tauri stars suggest that binary 
properties evolve as protostars are assembled out of their natal cores 
(e.g. Duch\^ene et al. 2004; Haisch et al. 2004; Connelley et al. 2008), 
so even results for Class II/III systems include the migration and 
dynamical interactions that occur after fragmentation; as we discuss 
further in Section 6, some of the properties of these slightly older 
systems could offer hints regarding very early evolution.

Our theoretical expectations for this young population are still 
highly uncertain. Several models have been proposed as the primary 
mechanism for multiple star formation (e.g. Bonnell 2001; Tohline 2002), 
with the three leading contenders being prompt fragmentation during an 
isothermal protostellar clump's initial freefall collapse, fission of a 
nonisothermal protostellar core after freefall collapse has ended, and 
disk fragmentation after the primary star has condensed and acquired a 
massive accretion disk. Fission seems to have been ruled out by 
hydrodynamical simulations, as a collapsing nonisothermal core will evolve 
on the Kelvin-Helmholtz timescale and angular momentum seems to be 
transported to the extremes of a non-axisymmetric core at a much faster 
pace (e.g. Durisen et al. 1986; Bonnell 1994). However, prompt 
fragmentation and disk fragmentation remain as viable explanations for 
different types of binary formation.

The most widely accepted model for the formation of wide ($>$100 AU) 
binary systems is by prompt fragmentation, during or just after the epoch 
where the prestellar core has become Jeans critical and begun free-fall 
collapse, but before the core has become nonisothermal and heating acts to 
oppose further collapse (as reviewed by Bodenheimer \& Burkert 2001). This 
process typically is modelled using smoothed-particle hydrodynamic (SPH) 
simulations (Bate 2000; Bate et al. 2002; Delgado-Donate et al. 2004); the 
most recent simulations of larger-scale star formation implicitly include 
this process by extending down to much smaller angular scales ($\sim$5 AU; 
Bate 2009a; Offner et al. 2009). After a protostellar core undergoes 
sufficient collapse to form a central protostar, the remaining envelope 
accretes into a circumstellar disk and from the disk onto the star. If the 
disk accumulates material from the envelope more quickly than mass can 
accrete onto the star, then it could grow more massive and violate the 
Toomre stability criterion (Toomre 1964), fragmenting to form a bound 
companion. This process has been modeled extensively for the formation of 
extrasolar planets (e.g., Boss 2001), but if there is sufficient material 
left in the disk and envelope, this bound companion would then accrete 
additional mass and grow into a stellar binary companion (e.g., Clarke 
2009). Observations suggest that the characteristic radius for a 
protostellar disk and for accretion onto it is 50--100 AU (Enoch et al. 
2009; Watson et al. 2007), so disk fragmentation could explain binary 
formation at small scales where prompt fragmentation is not feasible. 
Neither model has yielded quantitative predictions to date, but as we 
discuss in Section 6, we can use the predicted trends from these models 
and the observed properties of binary systems to infer some basic 
conclusions for multiple star formation.

One specific topic of recent interest is the formation of substellar 
binary companions. Over the past fifteen years, radial velocity surveys 
have discovered many short-period stellar companions and exoplanets, but 
relatively few companions with masses of $\sim$10--80 $M_{\rm Jup}$ (e.g., 
Marcy \& Butler 2000; Grether \& Lineweaver 2006), a gap known as the 
``brown dwarf desert''. Coronagraphic imaging surveys for wide companions 
have suggested that substellar companions might not be unusually rare, but 
instead could have a frequency consistent with an extension of the binary 
mass ratio function (e.g. Metchev \& Hillenbrand 2009). However, neither 
survey technique has been able to study the 5--50 AU regime, a separation 
range which represents the peak of the binary separation distribution for 
solar-type binaries, as well as the giant planet regime for our own solar 
system. Substellar companions bridge the mass range between binaries and 
exoplanets, so a census in this unexplored regime could indicate whether 
the substellar companion mass function and separation distribution more 
closely resemble the stellar or planetary cases. This census would also 
reveal the origin of the radial velocity (RV) brown dwarf desert; while it 
is possible that substellar companions never form at all, the paucity at 
small separations could also be traced to secondary effects like 
inefficient migration.

In this paper, we present a high-resolution imaging survey of the 
Taurus-Auriga star-forming region. Our survey uses adaptive optics and 
aperture masking interferometry with AO to achieve unprecedented angular 
resolution and depth, yielding a more complete view of the primordial 
multiple star population and the so-called brown dwarf desert. In Section 
2, we describe our survey sample, and in Section 3, we summarize the 
observations and our data analysis techniques. We summarize our new 
observational results for Taurus-Auriga and place them in the context of 
past surveys in Section 4, and then in Section 5, we characterize the 
binary properties for solar-type stars. Finally, in Section 6, we use 
these results to infer the processes and time- and length-scales of 
multiple star formation.

\section{Survey Sample}

The member census of Taurus-Auriga has been assembled gradually over the 
past several decades. The extremely low stellar density and variable 
extinction make it difficult and expensive to survey the association, 
especially away from the central cores. The wide range of evolutionary 
stages (from Class 0 protostars to Class III diskless stars) also result 
in a wide range in member properties, requiring numerous observing 
techniques to achieve completeness. We compiled a then-current (though 
still incomplete) stellar census in two previous works (Kraus \& 
Hillenbrand 2007a, 2008), and have used that census as the basis for our 
aperture masking sample. Many additional members have been identified by 
Luhman et al. (2006, 2009), Scelsi et al. (2008), and Rebull et al. 
(2010), but most are very low-mass stars or brown dwarfs that fall below 
the mass range of our survey.

Our original census selected every known member that had been shown to 
have at least one signature of youth (i.e. infrared excess, accretion, 
typical lithium abundance for 1--3 Myr old stars, or low surface gravity). 
We also explicitly required every member to have a known spectral type, so 
that we could infer a mass and study the mass-dependence of measured 
properties. This requirement rejected most of the very young Class 0/I 
sources, leaving only the more evolved Class II/III sources. The selection 
of our observed sample was subject to several biases. Natural guide star 
adaptive optics (AO) observations can be conducted only with a guide star 
that is optically bright ($R\la$15, with a brighter limit under marginal 
observing conditions). This requirement yields an effective joint limit in 
mass and extinction. The AO correction is also compromised for binary 
pairs with similar brightness ($\Delta$$R\la$1--2) and moderately close 
separations ($\sim$1--4\arcsec) since both sources are imaged on the 
wavefront sensor. Finally, we are unable to use aperture masking for the 
components of known binaries with separations of $\sim$0.4-1.0\arcsec\, 
because they are close enough for their interferograms to overlap, but too 
widely separated for the power spectrum to yield unaliased measurements.

In Table 1, we list all of the Taurus members that passed a preliminary 
spectral type cut (G0$\le$SpT$\le$M4, or 2.5$\ga$$M$$\ga$0.25 $M_{\sun}$ 
according to the methods described in Section 3.3) and have optical/NIR 
fluxes which are not dominated by scattered light (i.e. obscured by a 
circumstellar envelope, as for Class 0/I sources, or an edge-on disk). The 
spectral type range was chosen to match the nominal upper end of the 
Taurus mass function (which has $<$5 known A-F stars) and to avoid strong 
incompleteness for stars which are too faint for AO observations (with 
SpT$\ga$M5). The goal for this sample selection was to identify a sample 
of low-mass stars that are analogs of field solar-type or early-M stars.

The stars in Table 1 are divided between the observed sample (82 targets), 
known binaries that we did not reobserve (37 targets), and the stars that 
we could not observe (many of which have other, less sensitive 
multiplicity observations available). This last group is comprised of all 
stars fainter than $R=15$ (16 targets) and the similar-flux, moderately 
wider binary pairs that have $\Delta$$R<$2 and separations $<$4\arcsec (5 
targets). For binary systems with fainter secondaries, we tried when 
possible to simultaneously observe both components; otherwise, we 
concentrated on the primary. We also observed 6 stars that passed our 
observational selection criteria but have spectral types of $<$G0 or 
$>$M4, two stars that appear to be nonmembers (HBC 352 and HBC 353; Kraus 
\& Hillenbrand 2009b), two 1--2\arcsec\, binary companions that were 
serendipitously observed in the same images as their primary (StHa 34 B 
and RW Aur B), and one Class I source that served as a test for our 
ability to distinguish companions from extended emission (HL Tau). These 
stars are not included in our statistical analysis since the vast majority 
of Taurus members in those categories could not be observed, but we list 
these stars in Table 1 and will report their results for completeness.

We ultimately omitted 42 known binary systems, which introduces a bias 
against the detection of additional binary components that would denote 
hierarchical triple systems; if the presence of a wide tertiary influences 
subsequent fragmentation, then our results might not reflect the total 
population of Taurus. There are also 12 systems with separations of 
1--4\arcsec\, for which we could observe only the primary. As we discuss 
further in Section 5.2, much of this incompleteness can be remedied by 
using Bayesian analysis to infer the parameters of the binary population, 
but the validity of this correction depends on the degree of independence 
between wide binary formation and the subsequent fragmentation of their 
components into close pairs.

We also could not observe three known or suspected edge-on disk systems: 
Haro 6-5B, HH 30, and V710 Tau C. Furthermore, edge-on disk hosts are more 
difficult to identify in membership surveys since they do not fall on the 
association's photometric sequence, so there could be additional Taurus 
members that remain undiscovered. These two biases lead to some 
incompleteness for our multiplicity census among all disk hosts, but since 
the disk inclination is a purely geometric effect, it should not influence 
our conclusions.

We can partially remedy the incompleteness for unobservable binary systems 
by adopting the results of previous survey programs. There is a long 
history of multiplicity programs studying Taurus-Auriga, starting with the 
lunar occultation and speckle surveys of the 1980s and 1990s, and leading 
up to modern day searches using speckle imaging as well as natural and 
laser guide star AO. Many of the previous surveys labored under different 
selection biases than our own (i.e. NIR flux limits or the presence of 
tip-tilt guide stars, instead of our optical flux limits), so they 
complement our own sample and allow for a more complete mass-limited 
sample. We did not re-observe the known binary systems in order to 
increase the survey efficiency, and we will adopt the previous detection 
limits for members we could not observe. In Section 4, we present a census 
of the known binary systems and of the best detection limits for all 
apparently single stars.

\section{Observations and Analysis}

\subsection{Observations}

The technique of non-redundant aperture masking has been well-established 
as a means of achieving the full diffraction limit of a single telescope 
(e.g. Nakajima et al. 1989; Tuthill et al. 2000, 2006). The core 
innovation of aperture masking is to resample the telescope's single 
aperture into a sparse interferometric array; this allows for data 
analysis using interferometric techniques (such as closure phase analysis) 
that calibrate out the phase errors that limit traditional astronomical 
imaging by inducing speckle noise.  As we described in K08, aperture 
masking observations can yield contrasts of $\Delta K\sim$6 at 
$\lambda$$/D$ and $\Delta K\sim$4 at 1/3 $\lambda$$/D$, and we used the 
technique to identify over a dozen binary companions that fall inside the 
detection limits of traditional imaging surveys. More detailed discussions 
of the benefits and limitations of aperture masking, as well as typical 
observing strategies, can be found in the first paper of this series (K08) 
and in Readhead et al. (1988), Nakajima et al. (1989), Tuthill et al. 
(2000, 2006), Lloyd et al. (2006), Martinache et al. (2007), and Ireland 
et al. (2008).

We observed our survey targets over the course of two observing runs at 
Keck (in November 2007 and December 2008) and one observing run at Palomar 
(in November 2007). All of our targets were observed with the facility AO 
imagers, Keck/NIRC2 (Matthews et al., in prep) and Palomar/PHARO (Hayward 
et al. 2001), which have aperture masks installed in cold filter wheels at 
or near the pupil stop. Most observations were conducted using the $K'$ or 
$K_s$ filters, but we observed the brightest targets in both $H$ and $L'$ 
in order to maximize the resolution (in $H$, for close stellar companions) 
and depth (in $L'$, since low-mass companions should be very red). In all 
cases, we used a 9-hole aperture mask that passes 11\% of the total 
incident flux through nine 1.1m subapertures. The choice maximizes the 
throughput, as the other option (an 18-hole mask) passes half as much 
incident flux and can be used only with narrowband filters (due to 
wavelength-dependent dispersion in broadband filters) that are $\sim$10\% 
as wide as the corresponding broadband filters.

A typical interferometric measurement requires the observation of one or 
more source-calibrator pairs. However, our sample included numerous 
targets with similar positions and brightnesses, so instead, we observed 
groups of science targets and inter-calibrated between them; we described 
our observing methodology in more detail in Kraus et al. (2008). We 
started our survey by including association nonmembers (selected from 
2MASS to have colors consistent with distant giants) as external 
calibrators. However, we abandoned this tactic partway through the survey 
because the nonmembers were resolved into binary systems as often as the 
Taurus members were, which defeated their purpose. We summarize the 
observations for each group of targets in Table 2.

The observing sequences for these groups were identical to those described 
in Kraus et al. (2008), with each target being observed in three visits 
that each consisting of eight exposures with individual exposure time of 
20s. The total integration time per target was 480s, and including 
acquisition and observation overheads, the total time required per target 
was $\sim$15 minutes. All observations in $H$ and $K'$ were taken without 
dithering, such that the interferograms for all targets were placed in the 
same location on the detector (which was chosen to be free of bad pixels). 
Our analysis of dithered observations shows that on some telescopes, 
including different dither positions degrades the calibration (for reasons 
that we are still investigating), so we decided that it was more important 
to achieve good calibration for our (relatively bright) targets than to 
attempt sky subtraction of a background that was typically negligible. The 
$L'$ sky background is much more significant (2500 counts s$^{-1}$), so we 
chose a minimal two-point dither pattern. In this case, there was no 
evidence that the dithers affected the data analysis.

The observing conditions varied; our November 2007 observing runs were 
plagued by poor seeing and clouds, but our December 2008 observations were 
conducted under good to excellent seeing (0.2-0.5\arcsec\, in $K$ band). 
We therefore decided to re-observe those targets from 2007 that had 
maximum sensitivities of $\Delta K \la 4$ at the separation bin where we 
typically achieve optimal sensitivity (80-160 mas); the requirement of 
$\Delta K \ga 4$ insures that typical 1 $M_{\sun}$ sample members will 
have detection limits near the bottom of the brown dwarf mass range 
($M_{lim} \la$20 $M_{Jup}$). We will report both sets of detection limits.

\subsection{Data Analysis and Detection Limits}

The data analysis follows almost the same prescription as in Kraus et al 
(2008), so we discuss here only a general background to the technique and 
differences from Kraus et al (2008). The data analysis takes three broad 
steps: basic image analysis (flat-fielding, bad pixel removal, dark 
subtraction), extraction and calibration of squared-visibility and 
closure-phase, and binary model fitting. Unless fitting to close, 
near-equal binaries, we fit only to closure-phase, as this is the quantity 
most robust to changes in the AO point spread function (PSF). We converted 
the on-chip PAs to on-sky PAs using the most recent NIRC2 position angle 
calibration by Ghez et al. (2008), and treated the conversion between 
different $K$ filters ($K'$ versus $K_s$) as negligible (e.g. Carpenter 
2001) compared to the intrinsic uncertainties in relative photometry for 
AO observations ($\sim$0.03-0.05 mag).

The final detection limits are found using a Monte-Carlo method that 
simulates 10,000 random closure-phase datasets of a point source with 
closure-phase errors and a covariances that match those of the calibrated 
target target data set. This routine then searches for the best fit for a 
companion in each randomized dataset. Over each annulus of projected 
separation from the primary star, the 99.9\% (3.3$\sigma$) confidence 
limit (listed in Table 4) is set to the contrast ratio where 99.9\% of the 
Monte-Carlo trials have no best binary fit with a companion brighter than 
this limit anywhere within the annulus. The validity of this technique was 
demonstrated empirically by there being no spurious detections in Kraus et 
al (2008) above this limit, despite there being several near the limit.

A major difference between the Upper Scorpius analysis of Kraus et al 
(2008) and the Taurus analysis in this paper relates to the on-sky 
position angle of the observations. The aperture-mask was always used in a 
vertical angle mode at Keck, meaning the camera coordinates were fixed 
with respect to the elevation axis of the telescope, rather than being 
fixed with respect to N and E. For declinations that differed 
significantly from the telescope latitude, this meant that the position 
angle of the baselines changed with time, synthesizing a larger field of 
view. However, for Taurus, the declination is similar to the latitude of 
Keck, so if a target was observed while only rising or only setting, there 
was no sky rotation for aperture synthesis. We did not observe any targets 
within $<$30 minutes of transit so that any apparent rotation of a 
companion during a single integration would be $<<\lambda$$/D$.

Previous papers on aperture-masking (e.g. Martinache et al. 2007, 2009; 
Kraus et al 2008) have shown extracted visibilities and closure-phases. We 
do not repeat this here, but note that it is difficult to plot the raw 
closure-phases because the data are represented as a discrete set of 
points on a 4-dimensional grid. Instead, in Figure 1, we show two 
representative power spectra (i.e., 2-dimensional maps of the 
visibilities) taken using the 9-hole aperture-mask at Keck. The 
non-redundant geometry of the aperture mask leads to a power spectrum made 
of individual peaks (also referred to as splodges) corresponding to the 
baselines sampled by the mask. CIDA-9A shows an elongated core of the 
interferogram, and power that decreases to the top-left and bottom-right. 
However, closure-phases are zero within errors, meaning that this star has 
no close companions with good limits. We can confidently assign the 
elongation to a symmetrical instrumental cause (windshake in this case).  
CIDA-10 clearly shows two cores in the interferogram, and a corresponding 
sinusoidal modulation of the power spectrum amplitude when compared with 
CIDA-9. An even wider binary system would still have the core of the 
companion within the interferogram, but would show modulation within a 
single peak in the power spectrum. To reveal such wide companions, 
individual splodges need to be subsampled. However, this additional 
analysis only led to the detection of one additional companion (2M04414565 
Aa+Ab).

 \begin{figure}
 \epsscale{1.0}
 \plotone{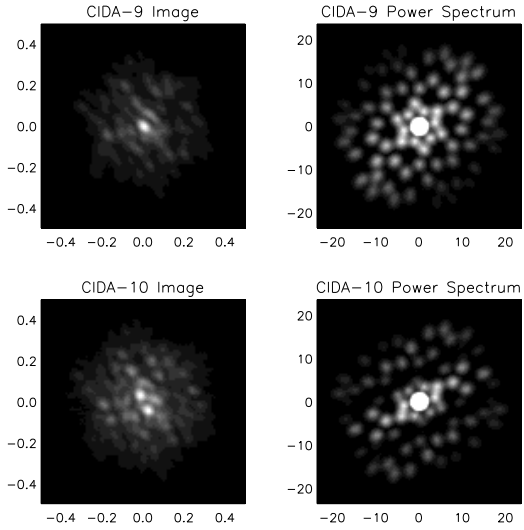}
 \caption{Interferograms (left) and power spectra (right) for the systems 
CIDA-9 A (no companion resolved with aperture-masking) and CIDA-10 AB. 
Units are in arcsec (for the images) and cycles per arcsec (for the power 
spectra), both in on-chip coordinates.}
 \end{figure}

\subsection{Stellar and Companion Properties}

Stellar properties can be difficult to estimate, particularly for 
young stars, since pre-main-sequence stellar evolutionary models are
not well-calibrated. The mass of a given sample could be
systematically uncertain by as much as 20\% (e.g. Hillenbrand \&
White 2004), and individual masses could be uncertain by factors of
2 or more due to unresolved multiplicity or the intrinsic
variability that young stars often display (from accretion or
rotational modulation of star spots). This suggests that any
prescription for determining stellar properties should be treated     
with caution.

We estimated the properties of our sample members using the method 
described in Kraus \& Hillenbrand (2007a), specifically adapted for stars 
of the median age of Taurus ($\sim$2 Myr; Kraus \& Hillenbrand 2009b). 
This procedure combines the 2 Myr isochrone of Baraffe et al. (1998) with 
the temperature scales of Schmidt-Kaler (1982) and Luhman et al. (2003) to 
directly convert spectral types to masses.  Relative properties (mass 
ratios $q$) for all binaries in our sample were calculated from the 
observed flux ratios ($\Delta K$ or $\Delta H$)  by combining these 
isochrones and temperature scales with the empirical NIR colors and 
bolometric corrections that we compiled in Kraus \& Hillenbrand (2007b). 
We note that this method assumes that both components are coeval (e.g. 
Kraus \& Hillenbrand 2009b) and have identical extinction; the latter 
assumption is untested for binary systems on the scale of our 
newly-discovered systems ($<$50 AU), and is known to fail for some wider 
systems (e.g. Connelley et al. 2008) including HL Tau/XZ Tau (Kenyon \& 
Hartmann 1995) and T Tau (Ratzka et al. 2009), as well as for systems like 
HV Tau AB-C where one component is seen only in scattered light (Duch\^ene 
et al. 2010). We also used these techniques to estimate masses for all of 
our sample members, which we list in Tables 1 and 5.

For all binary systems, we have adopted the previously-measured 
(unresolved) spectral type for the brightest component and inferred its 
properties from that spectral type. This should be a robust assumption 
since equal-flux binary components will have similar spectral types and 
significantly fainter components would not have contributed significant 
flux to the original discovery spectrum. We adopted a characteristic 
distance for all Taurus members of 145$\pm$15 pc. Recent high-precision 
parallax measurements with the VLBA (Loinard et al. 2007; Torres et al. 
2007, 2009) and from binary orbit fitting (Boden et al. 2007) suggest that 
there might be a distance gradient of 165-125 pc in the east-west 
direction, especially given the consistent distances of neighboring stars 
V773 Tau and Hubble 4 (136.2$\pm$3.7 pc versus 132.5$\pm$0.6 pc; Boden et 
al. 2007; Torres et al. 2007). However, a more detailed estimate of 
individual distances should be postponed until the overall structure of 
Taurus is better sampled.

Finally, for some of our sample members, the sensitivity limits of our 
survey extend to the bottom of the brown dwarf mass range and could 
potentially encompass the top of the planetary mass range. However, mass 
estimates for young giant planets are completely uncalibrated and there 
are ongoing debates regarding their peak and typical luminosities. The 
models of Baraffe et al. (2003) imply that a survey sensitive to $K\sim$16 
could detect 3 $M_{Jup}$ planets at the distance and age of Taurus. 
However, more detailed models of planet formation by Marley et al. (2007) 
and Fortney et al. (2008) suggest that accretion shocks can dispel much of 
the initial energy, leading to lower internal entropy and initial 
temperatures. At the typical age of Taurus members ($\sim$1--2 Myr), the 
luminosity of a planet could be 1--2 orders of magnitude lower than 
previously predicted. We can not currently resolve this controversy, so we 
only note that our limits on the presence of massive planets should be 
considered with caution.

\section{New Companions and the Multiple Stellar Population in 
Taurus-Auriga}

Our aperture masking observations are sensitive to near equal-flux 
companions at separations of $>$10 mas and can detect very faint 
companions ($\Delta$$m\sim$5.0--6.5 mag) at separations of $\ga$40 mas. 
The outer working angle for aperture masking, 360 mas, is set by the 
smallest baseline between subapertures. Companions outside this limit can 
still be identified, but masking observations lose sensitivity and are 
quickly surpassed in sensitivity by conventional AO imaging.  We 
conservatively estimate that visual inspection of the raw interferogram 
images (e.g., Figure 1, left) would have revealed any binary companions 
with separations of $\sim$0.3--2\arcsec\, and flux ratios 
$\Delta$$m\la$2.5, but none were found. In the vast majority of cases, 
this regime of parameter space has already been sampled by previous 
observations.

We list our newly-identified binary companions and the associated 
detection limits for all stars in Tables 3 and 4; we found a total of 16 
new companions among the 82 young stars we observed from our statistical 
sample. We found no additional companions to the other 10 stars that we 
observed in our campaign, but do not include in our statistical analysis. 
We also show the binary companions and observed detection limits for our 
sample in Figure 2, where we plot the flux ratio $\Delta$$m$ as a function 
of projected angular separation. In addition to the companions listed, 
there may be more companions below our detection limit that can 
nonetheless be confirmed by our data. For example, some of the points used 
to fit the orbit of GJ~802 in Ireland et al. (2008) would have fallen 
below our 99.9\% confidence limit here. Most notably, the transition disk 
system UX~Tau had a detection at 6.28 mags contrast and at a separation of 
65 milli-arcsec that was above the 99.9\% threshold by 0.03 mags. Its 
close proximity to the limits and potentially planetary nature suggest 
that we should treat it with caution until we can confirm it, so we do not 
yet include it in our analysis. We attempted to re-confirm the companion 
with deep L'-band aperture masking in 2009; based on a preliminary 
analysis, the data were good enough to detect the candidate companion if 
its color were $K' - L' > 0.5$. We are also in the process of making and 
analyzing follow-up observations of our binary detections. One object, 
LkCa 4 B, was not detected in the follow-up observations. At this point, 
we can not rule out either source variability or a yet unidentified 
systematic in data taken under poor seeing conditions, so we do not 
include it in our analysis either.

As we described in Sections 1 and 2, Taurus-Auriga has been the target of 
numerous multiplicity surveys over the past two decades. Though our 
results represent a significant leap forward, our newly-discovered binary 
systems still comprise only a significant minority of all known systems in 
Taurus. For separations of $<$4\arcsec, previous surveys have discovered 
57 additional binary companions to Taurus members with spectral types of 
G0--M4, some of which combine to form high-order multiple systems. 
Also, as we have described in past surveys, many of our observed targets 
are not truly independent systems but instead are bound into wide binary 
pairs with separations as wide as 30\arcsec (Kraus \& Hillenbrand 2008, 
2009a); our statistical sample includes 17 such pairs.

In Table 5, we summarize the properties of all binary pairs that have 
primary stars with $M=$0.25--2.5 $M_{\sun}$. This sample illustrates the 
wide variety of possible outcomes in multiple star formation, with the 
most highly multiple system (V955 Tau) containing at least six components. 
In Table 6, we list the corresponding detection limits from our survey and 
from the literature for all of the apparently single stars, as well as the 
limits for additional companions in all of the known multiple systems. We 
summarize the observed properties and total detection limits for our 
statistical sample in Figure 3, where we plot the mass ratio and companion 
mass as a function of projected physical separation.

As we show in Table 5 and Figure 3, several Taurus members in our sample 
have companions with apparently substellar masses, including some that 
fall well below the stellar regime ($M\la$50 $M_{Jup}$). These masses were 
determined only from the companion's flux ratio with respect to the 
primary, so some could be biased by systematic effects such as 
circumstellar disk excesses, accretion-based stellar variability, or 
differential extinction. Some binary companions, such as HV Tau C, HL Tau, 
and V710 Tau C, are even obscured by circumstellar envelopes or edge-on 
disks that completely block the star along our line of sight; they can 
only be seen in scattered light, and thus appear underluminous by many 
magnitudes. However, several of these companions are very likely to be 
substellar since they have known spectral types (e.g., 2M04414565 B) or 
are found in systems with no evidence that a disk is present (e.g., LkCa 5 
B, DI Tau B; Rebull et al. 2010) or at separations of $\la$50 AU (Haro 
6-37 Ab, 2M04080782 B) where binarity seems to prohibit formation of 
circumstellar disks (as will be shown in Kraus et al., in prep).

If we assume that this subset of companions is not biased by large 
systematic effects, then at least 5/129 or $>$3.9$^{+2.6}_{-1.2}$\% of the 
targets in our observed sample have a substellar companion with a 
separation of 5--5000 AU. This lower limit is similar to the 
completeness-corrected frequency of 3.2$^{+3.1}_{-2.7}$\% reported by 
Metchev \& Hillenbrand (2009) for a slightly narrower range of separations 
(28--1590 AU) and companion masses (12--72 $M_{Jup}$). However, the 
substellar companion frequency for Taurus may prove to be much higher if 
some of the wider companions (JH 112 Ab, JH 223 B, and StHa 34 B) are 
shown to be substellar or if some of the targets which were not amenable 
to masking observations host additional companions. We therefore suggest 
that the frequency could be higher by as much as a factor of $\sim$2 over 
this separation range. In fact, at least one additional sample member 
hosts a confirmed substellar companion (DH Tau B; Itoh et al. 2005), but 
the companion was not included in our statistical sample since the 
discovery survey did not report its null detections or detection limits.

Some of our observations were sensitive to even lower masses, with a 
handful reaching deep into the planetary-mass regime ($\sim$5--7 
$M_{Jup}$). Our survey of Upper Sco (Kraus et al. 2008) reached the 
planetary-mass range for many targets, allowing us to place constraints on 
the properties of the exoplanet population. However, most of the Taurus 
targets with the best contrast limits (i.e., which were observed in 
periods of the best seeing) are higher-mass stars, for which the same 
contrast limit could detect only higher-mass companions. Only 15 targets 
have contrast limits deep enough to detection a 10 $M_{Jup}$ planet at 10 
AU, and over half have detection limits of 15 $M_{Jup}$ or higher, so we 
have not repeated that analysis.

Finally, it is noteworthy that the young binary system HBC 427 already has 
been identified as a single-line spectroscopic binary by Mathieu et al. 
(1989). The companion that we identified has a projected separation of 32 
mas ($\sim$4.6 AU) and an apparent mass ratio of $\sim$0.7 (indicating a 
total system mass of $\sim$1.4 $M_{\sun}$). Massarotti et al. (2005) 
calculated an orbital solution with a period of $\sim$7 yr and an 
eccentricity of $e\sim$0.47; for a system mass of $\sim$1.4 $M_{\sun}$, 
their orbit places the semimajor axis and apastron distance at $\sim$4 AU 
and $\sim$6 AU, respectively. We therefore conclude that the companion we 
resolved is the unseen spectroscopic companion, and since the full RV 
curve has been determined already, this system presents an excellent 
prospect for a precise dynamical mass in the near future. This analysis 
has already been pursued by Steffen et al. (2001) using astrometry from 
the HST Fine Guidance Sensors, but an extrapolation of their orbital 
solution yields an inconsistent prediction for the companion position in 
our observations, so an updated solution seems to be required.

\begin{figure*}
 \epsscale{1.0}
 \plotone{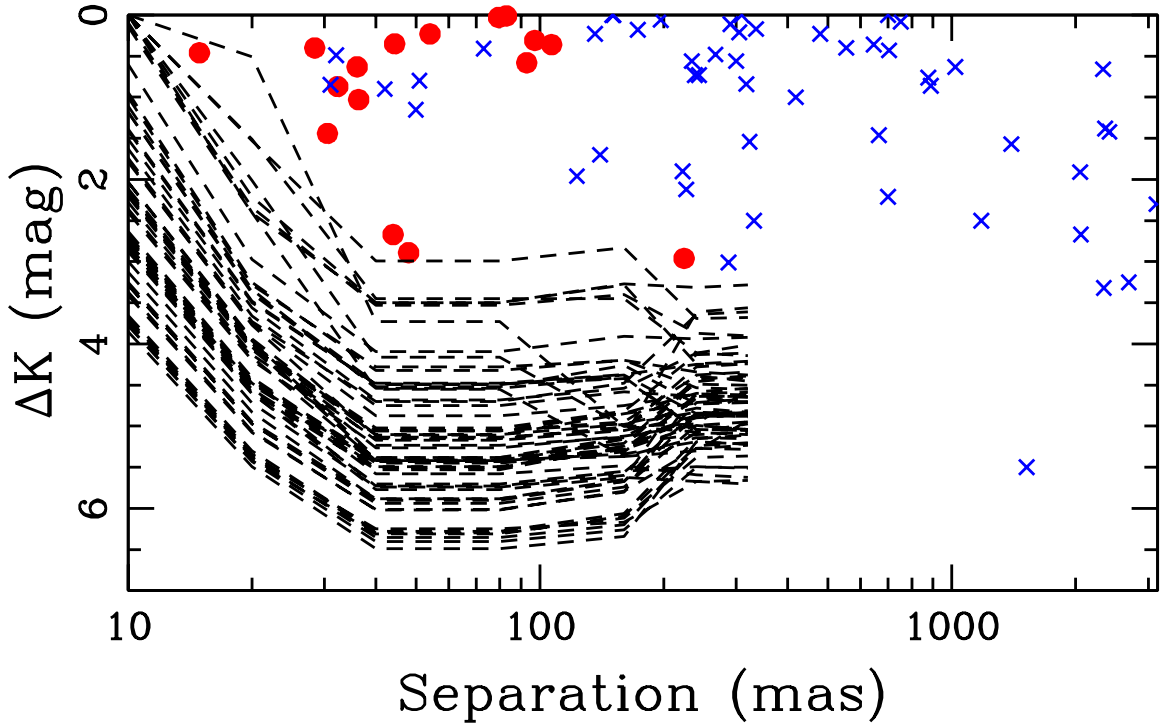}
 \caption{Detections and detection limits for our survey of Taurus Auriga. 
We plot the flux ratio $\Delta$$m$ (in magnitudes) as a function of 
projected separation (in mas) for each of our newly discovered companions 
(red filled circles) and the known companions to our sample members (blue 
crosses). We also show the detection limits for all apparently single 
stars in our sample (black dashed lines). Most companions fall well above 
the detection limits, but some companions could be substellar if their low 
luminosity is not a result of a systematic effect (such the presence of 
obscurative circumstellar material along the line of sight, as for HV Tau 
C or FV Tau B).}
 \end{figure*}

 \begin{figure*}
 \epsscale{1.0}
 \plotone{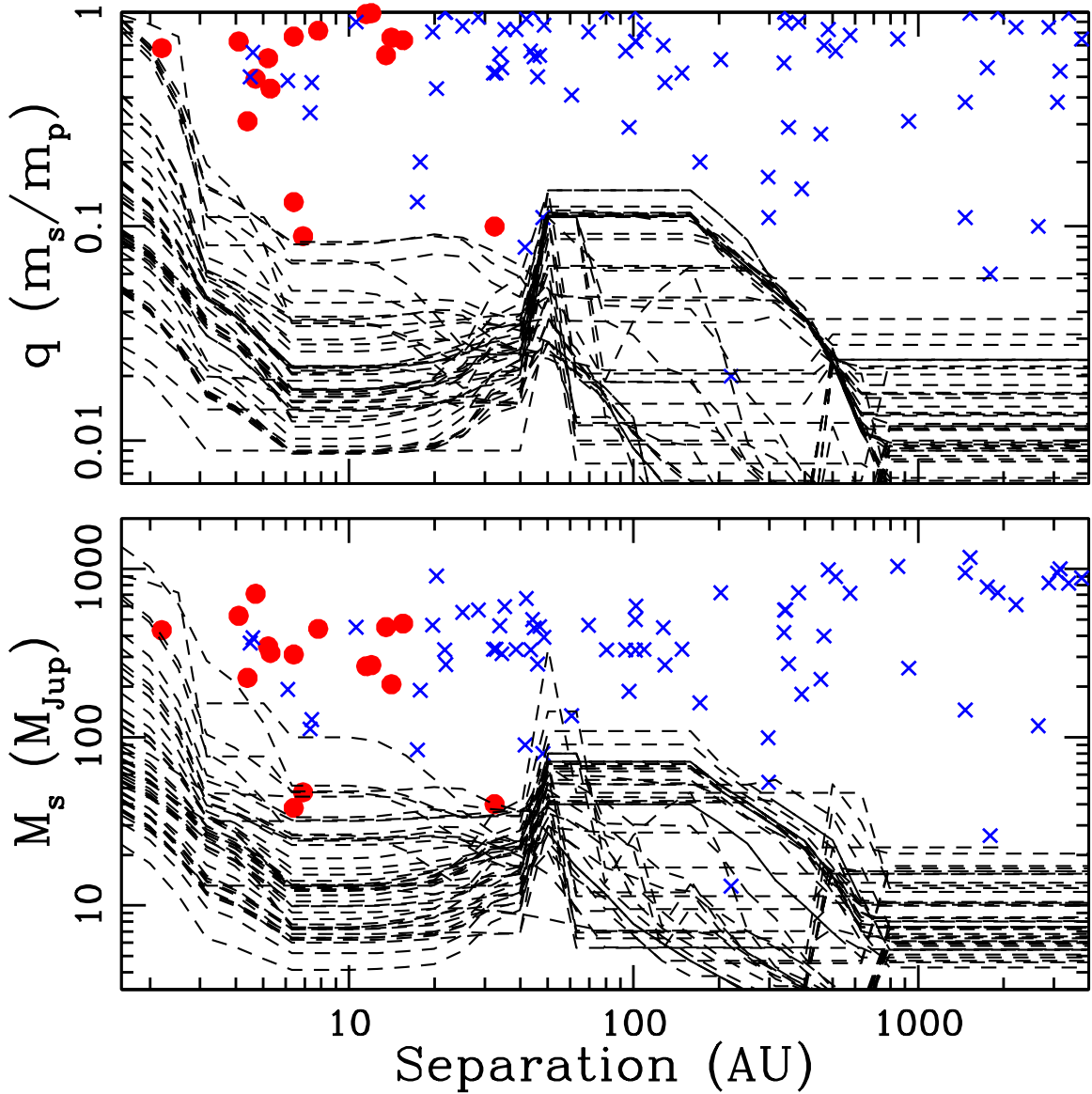}
 \caption{Detections and detection limits for our statistical sample, 
encompassing all of our observations as well as the detection limits 
adopted from the literature. The limits have been converted into mass 
ratios ($q=m_s/m_p$) and companion masses (in $M_{Jup}$). Symbols are the 
same as for Figure 2. Spikes in the typical detection limits can be seen 
at separations of $\sim$50 AU and $\sim$500 AU; these result from gaps 
where the outer working angle of one observing technique does not quite 
reach the inner working angle of another technique. Our Bayesian analysis 
(Section 5.2) naturally accounts for these narrow regions.}
 \end{figure*}

\section{The Mass-Dependent Properties of Young Multiple Systems}

Our new, significantly more complete binary census of Taurus offers a 
unique opportunity to study the primordial outcome of multiple star 
formation. In the following subsections, we approach this question from 
several angles. First, we construct histograms of the separation 
distribution and mass ratio distribution in order to determine what 
functional forms seem most appropriate for their description. Next, we use 
Bayesian analysis techniques to estimate the relevant scale parameters for 
those functional forms. Finally, we address the supposed ubiquity of 
multiple star formation by directly counting the number of apparently 
single stars in Taurus-Auriga.

\subsection{Observed Distributions}

The current paradigm for field binary properties was established by DM91, 
who conducted a volume-limited multiplicity survey of solar-type stars 
with spectral types F7-G9. They found a separation distribution which is 
apparently unimodal and log-normal, with a mean semimajor axis of $\sim$30 
AU and a standard deviation of $\sim$1 dex. They also found a mass ratio 
distribution that is peaked at low masses ($q\sim$0.3) and has few 
similar-mass companions, though their survey was not sensitive to most 
substellar companions and relied on significant completeness corrections 
for low-mass stellar companions. Finally, they found that $\sim$60\% of 
solar-type stars have at least one binary companion. The frequency and 
properties of binary systems appear to depend on their mass (e.g. FM92; 
Bouy et al. 2003; Burgasser et al. 2003; Close et al. 2003), but the mass 
range of DM91 is well-matched to the median mass for the upper half of our 
sample.

Subsequent surveys of young stars have not observed the same features as 
in DM91, especially for less dynamically evolved populations. Our survey 
of wide multiplicity in Taurus and Upper Sco suggests that the separation 
distribution for solar-type stars is actually log-flat, with more wide 
binary companions than are seen in the field (Kraus \& Hillenbrand 2009a). 
We also found in our aperture-masking survey of Upper Sco (Kraus et al. 
2008) that the mass ratio distribution of solar-type binaries might be 
much less biased toward low-mass companions, with the most likely 
distribution being linearly-flat such that all companion masses are 
equally probable. Most surveys of young stars in these regions (e.g., Ghez 
et al. 1993, Leinart et al. 1993; Simon et al. 1995; Kouwenhoven et al. 
2007; Kraus et al. 2008) find a significantly higher binary frequency than 
in the field, such that the binary frequency of solar-type stars in Taurus 
might approach 100\%.

These discrepancies between the field and young star-forming regions show 
that we can not assume the binary properties in our sample match those in 
the field, so we begin our analysis with the simplest non-parametric 
analysis: plotting histograms of the binary properties. To this end, we 
plot the separation distribution for a range of mass ratios where our 
survey is nearly complete, then we plot the mass ratio distribution for a 
range of separations where our survey is nearly complete. We must 
implicitly assume that the separation distribution and mass ratio 
distribution are not correlated, but our surveys of multiplicity at small 
and large separations (e.g. Kraus et al. 2008 versus Kraus \& Hillenbrand 
2009a) show no such correlation in other samples of similar size, so this 
assumption should be robust for our new Taurus sample.

In Figure 4, we show the observed separation distribution for our full 
sample and for two subsets of primary masses. Each bin of mass and 
separation represents the frequency of companions among all sample members 
for which we could have detected binary companions with mass ratios of 
$q\sim$0.1, so the number counts vary between bins and are generally 
higher at larger separations (where it is easier to achieve deep detection 
limits). We also treat close binary pairs as a single combined mass for 
the purposes of tertiary companion detection, so many of the close 
binaries from the low-mass subsample range are represented in the 
wide-separation bins of the high-mass subsample. In each case, we also 
plot the separation distribution observed by DM91 for both their total 
binary frequency ($\sim$39\%) and our observed total binary frequency 
(63--76\%) across this separation range.

The overall mass-dependent trends match our expectations from previous 
surveys of young stars (Kraus et al. 2008; Kraus \& Hillenbrand 2009a). 
The separation distribution for approximately solar-mass stars (0.7-2.5 
$M_{\sun}$) appears log-flat over separations of 5--5000 AU, similar to 
our results for Upper Sco (K08; Kraus \& Hillenbrand 2009a). For 
lower-mass stars (0.25-0.70 $M_{\sun}$), wide ($\ga$200 AU) binary 
companions are less common, though binary companions at smaller 
separations remain common. It is unclear whether the separation 
distribution is better modeled as a log-normal function (as for the field) 
where the mean and standard deviation decline with mass, or as a log-flat 
function with a mass-dependent outer cutoff. As we describe in the next 
subsection, modeling the function as log-flat leads to trivial 
conclusions, so we will emphasize the log-normal distribution in our 
subsequent analysis.

In Figure 5, we show the mass ratio distribution for all binary companions 
with projected separations of $>$50 mas ($\ga$15 AU) and mass ratios 
$>$0.1, again for our entire sample and for two ranges of primary mass. 
Each bin of mass and mass ratio represents the number of companions among 
all of the sample members for which we could have detection binary 
companions with mass ratios of $q\sim$0.1 at separations of $\sim$100 mas. 
We can not present a frequency for each bin, unlike for Figure 4, because 
the effective ``primary mass'' changes as a function of separation for 
systems with multiple components. Any attempt to correct for including 
partial separation ranges would require untested assumptions about the 
separation distribution, so we prefer to compromise by dealing with number 
counts instead of frequencies, then addressing the more comprehensive 
population statistics in our Bayesian analysis.

The full mass ratio distribution is close to flat, but with a moderate 
excess of similar-mass companions. This distribution is a stark contrast 
to the results of DM91, who found few similar-mass companions. However, it 
is much more consistent with previous surveys of young stars, which found 
a distribution that was close to linearly flat for solar-type stars and an 
increasing tendency for similar-mass companions at $\la$0.3 $M_{\sun}$. 
The mass ratio distribution for 0.25-0.7 $M_{\sun}$ stars has more 
similar-mass companions than low-mass companions, but as we will show more 
clearly in the next subsections, the difference is only marginally 
significant. We therefore will follow the lead of previous Bayesian 
analysis implementations and will treat the mass ratio distribution as a 
power law with an unconstrained exponent.

 \begin{figure}
 \epsscale{1.0}
 \plotone{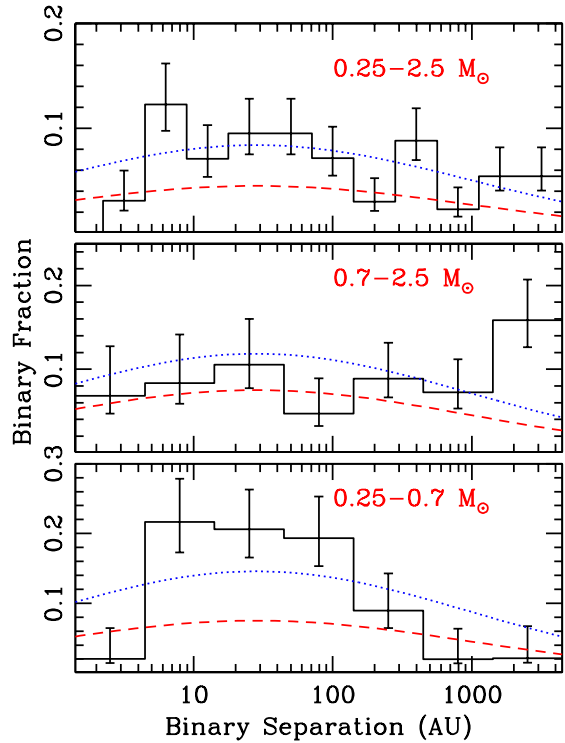}
 \caption{Separation distribution for our full sample (top) and for two 
subsets of primary mass. For each bin of primary mass and projected 
separation, we plot the companion fraction for all stars which were surveyed 
to a sensitivity of $q\sim$0.1 or better. The red dashed line denotes the 
separation distribution observed by DM91 as normalized to their companion 
fraction (39\% in this separation range), while the blue dotted line shows 
the same separation distribution renormalized to match the total companion 
fraction of our sample in that mass range (73\% for all stars, 62\% for 
the high-mass subsample, and 77\% for the low-mass subsample).}
 \end{figure}

 \begin{figure}
 \epsscale{1.0}
 \plotone{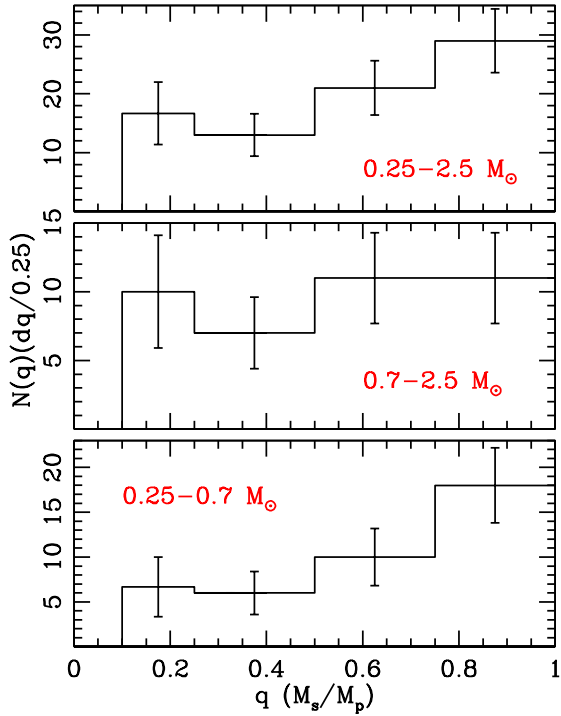}
 \caption{Mass ratio distribution for our full sample (top) and for two 
subsets of primary mass. For each bin of primary mass and mass ratio, we plot 
the number of binary companions among all systems with projected separation 
$>$50 mas. As we describe in the text, we must consider number counts 
instead of frequencies because systems with multiple components have a 
different ``primary mass'' for the close and wide companion, rather than a 
single definition that spans all separations. The low-$q$ bin only extends to 
$q=0.1$, so we have cast the y-axis as the number of companions per 
$\Delta$$q=$0.25, then multiplied the number of observed companions with 
$0.10<q<0.25$ by a factor of 5/3 in order to match this definition.}
 \end{figure}

\subsection{Bayesian Analysis}

Binary population statistics are traditionally presented in terms of 
histograms of companion frequency versus separation or mass ratio, where 
the data is presented only for a range where the survey is complete (e.g., 
Figures 4 and 5). The analytic form of the preferred model is then fit to 
these histograms in order to infer the population properties. As we showed 
in the previous subsection, this non-parametric approach is required in 
cases where the functional forms of the parameter distributions are 
unknown. However, once a functional form can be prescribed, then a better 
solution for working with heterogeneous data is to adopt a Bayesian 
approach, where the scale parameters of the model are assigned a prior PDF 
and that PDF is modified by each observation. This method exploits Bayes' 
theorem:

\begin{equation}
 P(\theta|O) \propto P(O|\theta) P(\theta)
\end{equation}

\noindent where $\theta$ represents the ``model'' (a set of scale   
parameters describing the functional form), $O$ represents the
observation, $P(\theta|O)$ is the posterior PDF for the model (as a   
function of its parameters) given the data, $P(O|\theta)$ is the
probability of obtaining an observation as a function of the model   
parameters, and $P(\theta)$ is the prior PDF for the model (again, as a
function of its parameters). In cases with multiple observations, the
posterior function for one observation is then used as the prior function
for the next observation.

Bayesian analysis techniques offer several compelling advantages over 
traditional techniques that produce histograms and fit probability density 
functions. The most notable distinction is that Bayesian analysis 
optimally exploits the available data, while implicitly avoiding any need 
for completeness corrections. Traditional histogram analysis requires the 
identification of a ``complete'' regime of parameter space where all 
observations are sensitive to the detection of companions; in some cases, 
this regime can be expanded by using a completeness correction for regimes 
of partial sensitivity. In contrast, our formulation of Bayesian analysis 
optimally exploits each observation by drawing value from regimes where a 
companion could be detected, but ignoring regimes where one could not. 
Bayesian analysis also avoids the uncertainties of binning, which can be 
significant if there are only enough observations to justify a small 
number of bins (e.g. Figure 5). Finally, Bayesian analysis provides a more 
direct measurement of physically meaningful parameters. Histograms measure 
the PDF of the population, and then that PDF must be fit with 
distributions in order to estimate the parameters that describe that 
population (such as the mean separation or the slope of the mass ratio 
distribution).  In contrast, Bayesian analysis directly yields the PDF for 
those parameters, bypassing the intermediate step. As we show below, this 
advantage can be helpful not just in showing a study's measurements of 
population properties, but also honestly presenting the limits on those 
measurements and the extent of ignorance. However, we must acknowledge a 
significant caveat. Bayesian analysis is only meaningful for assumed 
functional forms of a population, and as we showed above, histograms must 
be inspected first to determine that a given functional form appears 
valid.

Allen (2007, hereafter A07) developed the relevant techniques for applying 
Bayesian statistics to multiple star populations, and we used his approach 
in our recent survey of multiplicity in the very low mass (VLM) regime 
(Kraus \& Hillenbrand 2010). This method describes the PDF for the binary 
population in terms of a companion frequency $F$, a power-law mass ratio 
distribution with exponent $\gamma$, and a log-normal separation 
distribution with mean $\overline{\log(\rho)}$ and standard deviation 
$\sigma$$_{\log(\rho)}$. Each of these parameters is given a prior, and 
then the observations modify this prior to yield the posterior PDF that 
carries our new constraints on the population. However, rather than using 
the conventional Bayesian approach where each observation serially 
modifies the prior, this method instead compiles a single ``window 
function'' $N_{obs}(q,\log(\rho))$ that describes the number of 
observations which are sensitive to discrete bins of separation 
$\log(\rho)$ and mass ratio $q$, plus a corresponding ``companion 
function'' $N_{comp}(q,\log(\rho))$ that describes the number of 
companions detected in each of those bins. The net effect is to treat each 
bin of parameter space ($\Delta q$,$\Delta \log(\rho)$) as an observation, 
then iterate through all bins so that they serially modify the prior PDF 
to yield a posterior PDF.

Given this assumed functional form that describes the population, for each 
set of model parameters the expected frequency of companions in a bin 
($\Delta q$,$\Delta \log(\rho)$) is given by a probability $R$ such that:

 \begin{eqnarray}
 &R&(\log(\rho),q|F,\overline{\log(\rho)},\sigma_{\log(\rho)},\gamma) 
\Delta \log(\rho) \Delta q = \nonumber \\
&&\frac{\gamma+1}{\sqrt{2\pi}\sigma} F q^{\gamma} 
\exp(-\frac{(\log(\rho)-\overline{\log(\rho)})^2}{2\sigma_{\log(\rho)}^2}) 
\Delta q \Delta \log(\rho) 
 \end{eqnarray}

\noindent For this probability $R$, the corresponding value of 
$P(O|\theta)$ for our observed total number of companions $N_{comp}$ and 
total number of observations $N_{obs}$ in that range of ($\Delta 
q$,$\Delta \log(\rho)$) is then given by the Poisson likelihood function:

 \begin{eqnarray}
 P(N_{comp},N_{obs}|F,\overline{\log(\rho)},\sigma_{\log(\rho)},\gamma) 
\propto R^{N_{comp}} \times e^{-R \times N_{obs}}
 \end{eqnarray}

\noindent We iterated our calculation over all mass ratios from 0 to 1 in 
steps of 0.01 and over all values of $\log(\rho)$ between 0.2 and 3.6 dex 
in steps of 0.1 dex, allowing each bin of ($\Delta q$,$\Delta \log(\rho)$) 
to serially modify the prior PDF and generate the posterior PDF.

As in our other work, we adopt several modifications to the formalism of 
A07. The most significant feature is to assume constant prior values for 
$F$ and $\sigma$$_{\log(\rho)}$, whereas the description in A07 suggests 
that he might have adopted priors proportional to $1/F$ and 
$1/\sigma$$_{\log(\rho)}$, respectively; we believe that the constant 
priors are more appropriate for an unbiased analysis with minimal initial 
assumptions. We chose to model the separation distribution in terms of 
observed projected separation rather than the underlying semimajor axis 
distribution. If the separations and eccentricities are uncorrelated, then 
the two distributions are related by a constant multiplier that depends on 
the eccentricity distribution (e.g. FM92), and we prefer to work with the 
observed quantity rather than an uncertain inferred quantity. We also 
omitted the flux-completeness correction used by A07 to compensate for the 
overluminosity of similar-brightness binaries. The discovery surveys for 
most of our sample members were spatially limited, not flux-limited, so 
binary systems were as likely to be detected as single stars. The 
detection limits of wavefront sensors in adaptive optics imaging are 
generally optically flux-limited, but we chose our mass cutoff to 
alleviate this problem and have invoked the results of previous surveys to 
further account for any remaining incompleteness (Section 5; Table 6; 
Figure 3). Finally, our population parameter $F$ is more formally treated 
as a companion frequency (the number of companions per primary star) 
rather than a binary frequency (the number of primary stars with at least 
one companion). This distinction did not matter for the sample analyzed in 
Allen (2007) since it included no high-order multiple systems, but our 
sample includes many systems where one primary has more than one 
companion, and all of these companions contribute to the overall companion 
frequency per primary star.

Previous surveys have shown that the frequency and properties of multiple 
systems depend on the system mass (e.g., DM91 vs FM92), so we have 
conducted this analysis for the entire sample and for two subsets of 
primary masses: 0.7-2.5 $M_{\sun}$ and 0.25-0.7 $M_{\sun}$. The division 
between these subsets is located at the same mass as for our survey of 
Upper Sco, which will allow us to directly compare the results of our two 
surveys. However, our Taurus sample spans a wider total range of mass than 
that of our Upper Sco sample (0.25--2.5 $M_{\sun}$ versus 0.4--1.7 
$M_{\sun}$), so any mass-dependent trends should be evaluated accordingly.

As we noted in the previous section, we also must consider an alternate 
separation distribution, which is a log-flat distribution with 
mass-dependent inner and outer cutoffs. However, we can only constrain one 
such cutoff across the separation range of our sample (the outer cutoff 
for lower-mass stars), so the results of Bayesian analysis for that model 
are trivially unconstrained. As we describe in the next subsection, we can 
use the results of our Bayesian analysis to estimate the 
completeness-corrected binary frequency for separations of 3--5000 AU, 
which allows us to directly estimate the companion frequency per decade of 
separation, and thus a relation between the total companion frequency and 
the interval spanned between the inner and outer cutoffs of a log-flat 
function.

\subsection{The Primordial Multiplicity of Solar-Type Stars}

Our Bayesian analysis yields a PDF for all possible ``models'' that is 
defined across four dimensions ($F$, $\gamma$, $\overline{\log(\rho)}$, 
and $\sigma_{\log(\rho)}$), so we can not present the full results in 
a two-dimensional medium. However, any uncorrelated parameters can be 
presented separately without discarding information. This independence 
allows us to present the results as a series of lower-dimensional 
surfaces, where the PDF is integrated across the uncorrelated parameters 
in order to flatten its dimensionality. We have found that our constraints 
on $\overline{\log(\rho)}$ and $\sigma$ are strongly correlated, while our 
constraints on $\gamma$ are not correlated with any other parameters, so 
we present our results in terms of two planes ($F$ versus 
$\overline{\log(\rho)}$ and $F$ versus $\sigma$) and one interval 
($\gamma$).

In Figure 6, we show our joint constraints on the companion frequency, 
mean separation, and standard deviation of the separation, inferred for 
the entire sample and then for the high-mass and low-mass subsamples. The 
observed frequency of companions($\sim$70\% in each mass range) placed a 
lower limit on the overall companion frequency.  However, since there 
could be a significant number of companions inside or outside the survey 
detection limits, significantly higher frequencies are allowed. Our 
results even allow for companion frequencies of $>$100\%, which would 
indicate a significant population of hierarchical multiple systems.

For the high-mass subsample, the nearly log-flat separation distribution 
yields a wide range of allowed values. There is no correlation between the 
most probable mean separation and the frequency, but the range of allowed 
mean separations is correlated; we have observed that the companion 
frequency is $\sim$70\% for separations of $\sim$3--5000 AU, so if the 
mean separation is not located at the logarithmic mean of this range 
($\sim$200 AU), then there must be additional companions beyond the inner 
or outer working angles of our survey. By similar reasoning, the standard 
deviation of the separation distribution is strongly correlated with the 
frequency. The nearly log-flat distribution for separations of 3-5000 AU 
indicates that additional companions (beyond those we observe) must be 
spread across a wide range of inner or outer separations, or else the 
distribution across our observed range would not appear flat.

For the low-mass subsample, the paucity of companions at separations of 
$\ga$200 AU clearly indicates a more restricted set of preferred models. 
If the companion frequency is significantly higher than the observed value 
of $\sim$70\%, then most of the additional companions must be found inside 
the inner working angle of our survey. As a result, a higher frequency is 
strongly correlated with a smaller mean and a larger standard deviation in 
the separation distribution. The relatively sharp outer limit in the 
binary population also weighs against significantly larger values of 
$\sigma_{\rho}$ (and thus higher frequencies) since an extended tail 
should not show such an abrupt decline.

In Figure 7, we show our confidence intervals for $\gamma$, again for the 
entire sample and for both subsamples. We have found that our constraints 
on $\gamma$ are not significantly correlated with the other parameters, a 
result of our survey's sensitivity to even very low mass ratios ($q<0.1$) 
across most of its separation range. There is also little evidence for a 
mass dependence in the mass ratio distribution. The estimated power law 
slope for the entire sample, $\gamma$=0.2$\pm$0.2, is consistent at the 
1$\sigma$ level with the values for the high-mass subsample 
($\gamma$=0.0$\pm$0.2) and the low-mass subsample ($\gamma$=0.4$\pm$0.2). 
However, studies of intermediate- to high-mass stars (2--10 $M_{\sun}$; 
Kouwenhoven et al. 2007) and very low-mass stars and brown dwarfs($\la$0.3 
$M_{\sun}$; Burgasser et al. 2006; Allen 2007; Kraus \& Hillenbrand 2010) 
demonstrate that a large scale trend does exist. The power-law slope for 
intermediate- to high-mass stars is negative ($\gamma \sim$-0.4; 
Kouwenhoven et al. 2007), while the slope becomes increasingly negative 
near and below the substellar boundary ($\gamma \sim$2--4; Kraus \& 
Hillenbrand 2010). We therefore suggest that either this entire mass range 
shows a gradual trend toward similar-mass companions at lower masses, or 
there are multiple processes that set the mass ratio distribution in 
different primary mass ranges. 

Interestingly, the DM91 mass function also appears to be have a negative 
slope. They did not report a power-law fit, so we refit the histogram for 
their completeness-corrected $q$ distribution with a power law. We found 
that the entire distribution ($0<q<1.1$) has a best-fit slope of $\gamma = 
-0.36 \pm 0.07$, albeit with a poor fit ($\chi_{\nu}=2.7$ with 9 degrees 
of freedom). If we omit the two lowest-mass bins (which consist largely of 
completeness correction) and only fit the remaining range ($0.2<q<1.1$), 
we find a much better fit ($\chi_{\nu}=0.7$ with 7 degrees of freedom) and 
a much steeper negative slope of $\gamma = -1.2 \pm 0.2$. The former value 
disagrees with our entire sample at $\sim 3\sigma$ and with our solar-type 
subsample at $\sim 2\sigma$, while the latter value disagrees at $\sim 
5\sigma$ and $\sim 4\sigma$, respectively. As we discuss below, this might 
reflect the application of too large a completeness correction for systems 
with low mass ratios, especially since updated surveys (e.g. Raghavhan et 
al. 2010) also report a shallower mass ratio distribution for field 
solar-type stars.

As we discussed in Section 1, the theoretical expectation is that binary 
systems with separations of $\ga$100 AU most likely form via freefall 
fragmentation during early collapse, while systems with separations of 
$\la$100 AU most likely form via fragmentation of the protostellar disk 
after the primary had ceased freefall collapse. These two processes occur 
at different times and should proceed in a very different fashion, so it 
seems plausible that they might produce binary companions with a different 
mass function. We tested for this difference by independently analyzing 
our results with the Bayesian formalism for the two separation ranges, 
then marginalizing the resulting PDFs to yield measurements and confidence 
intervals for $\gamma$; this yields PDFs like those shown for our mass 
subsamples in Figure 7. Contrary to our expectation, we find that there is 
no evidence for a different mass function at large separations than at 
small separations. The best-fit slope of the mass function is $\gamma = 
0.22 \pm 0.22$ at separations $\la$100 AU and $\gamma = 0.08 \pm 0.20$ at 
separations $\ga$100 AU, and hence the two mass functions are consistent 
to within 0.5$\sigma$.

Finally, we must conclude by explicitly noting the degeneracies in our 
parameter constraints that result from our survey's inner and outer 
working angles. The degeneracy due to the inner working angle will be 
broken by results from radial velocity surveys, including ongoing programs 
by Prato et al. (2008), Nguyen et al. (2009), and White et al. (in 
preparation). Breaking the degeneracy for extremely wide binary systems 
will be more difficult since binary systems must be distinguished from 
chance alignments of unbound stars (e.g. Kraus \& Hillenbrand 2008), but 
we are developing statistical tools for measuring and subtracting this 
contamination. In the meantime, we can avoid these degeneracies by 
forward-modeling from our four-dimensional PDF back into the range of 
parameter space where our survey is mostly complete. This extrapolation is 
effectively a completeness correction that integrates the correction over 
all possible models, weighted by the probability of each model. To this 
end, we have integrated over the entire four-dimensional PDF of each mass 
subsample to extrapolate the binary frequency at separations of 3-5000 AU 
and spanning all mass ratios. We find that the total companion frequency 
in this range of parameter space is 64$^{+11}_{-9}$\% for 0.7-2.5 
$M_{\sun}$ stars and 79$^{+12}_{-11}$\% for 0.25-0.7 $M_{\sun}$ stars.

 \begin{figure*}
 \epsscale{1.0}
 \plotone{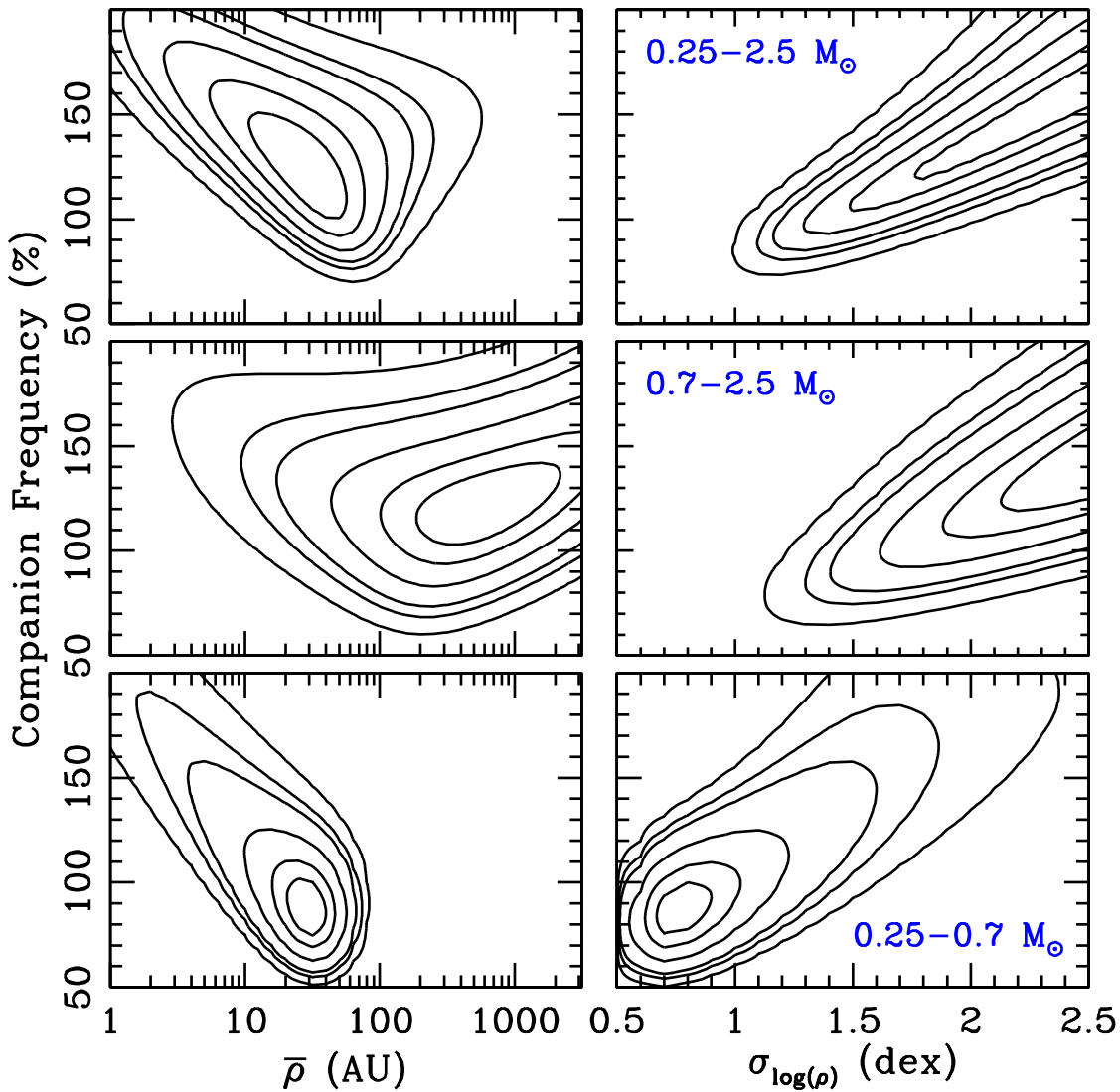}
 \caption{Joint constraints on the companion frequency versus the mean 
(left) and standard deviation (right) of the separation distribution, as 
computed for the entire sample (top), only stars with $M>$0.7 $M_{\sun}$ 
(middle), and only stars with $M<$0.7 $M_{\sun}$ (bottom). Contours are 
drawn to enclose 25\%, 50\%, 75\%, 90\%, 95\%, and 99\% of the total 
probability. The apparently log-flat separation distribution for 
solar-type binaries is indicated by the wide range of possible mean 
separations and the strong tendency for large values of the standard 
deviation. For lower-mass binaries, the paucity of wide binary companions 
indicates that small values of the mean separation are preferred, though 
the unknown form of the distribution at separations $\la$3 AU yields a 
degeneracy between the companion frequency and mean separation. In both 
cases, the observed frequency of binary companions places a strong lower 
limit on the possible frequency ($\ga$60-70\%).}
 \end{figure*}

 \begin{figure}
 \epsscale{1.0}
 \plotone{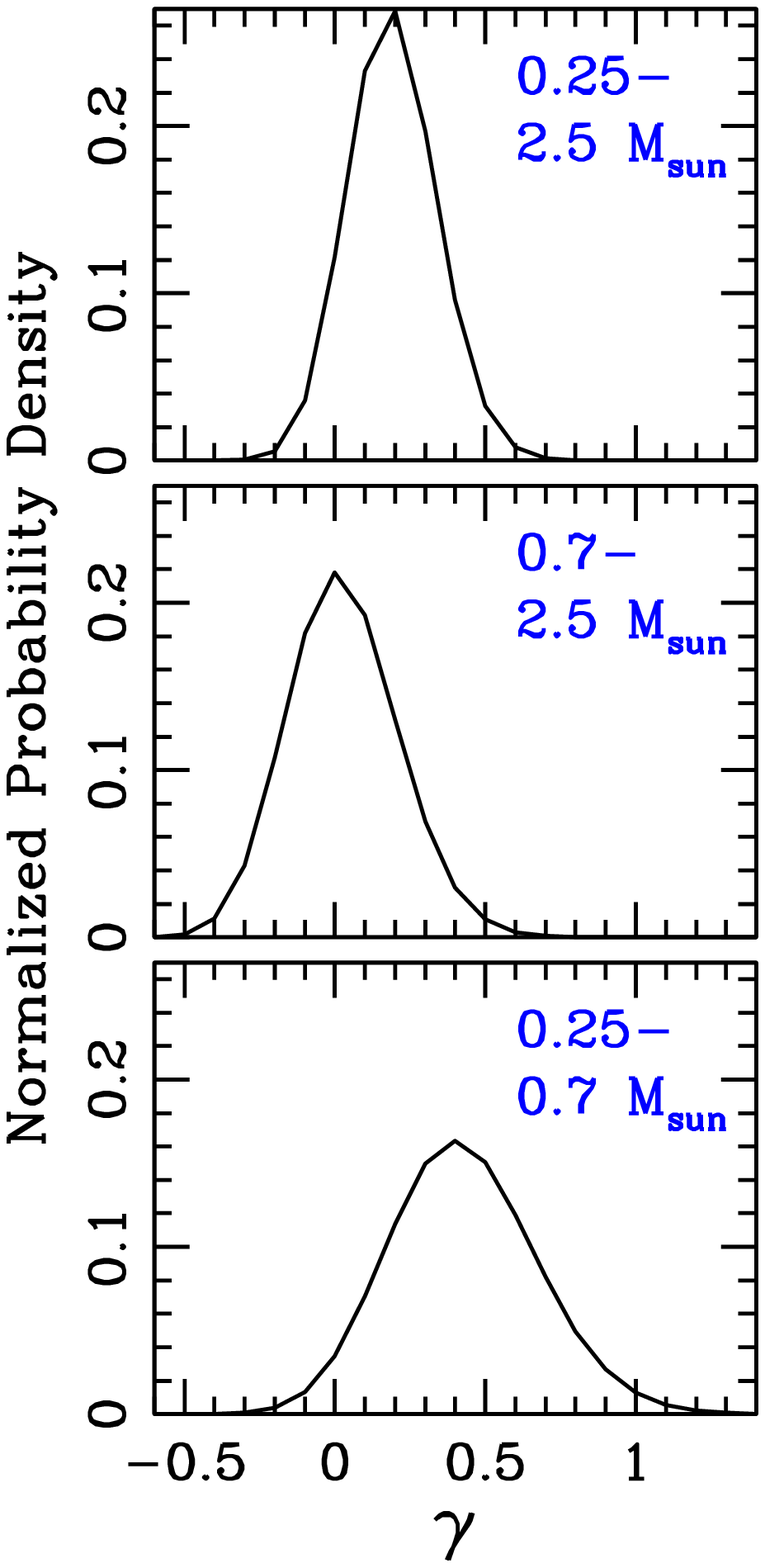}
 \caption{Confidence intervals for $\gamma$, the power law exponent of the 
companion mass ratio distribution. As for Figure 6, we show our results 
for the entire sample, only $M>$0.7 $M_{\sun}$, and only $M<$0.7 
$M_{\sun}$. We find that all three cases yield the same value to within 
the uncertainties, and are typically consistent with a linear-flat case 
($\gamma = 0$).}
 \end{figure}

\subsection{The Frequency of Single Stars}

It has been suggested that binary formation might be required for the vast 
majority of protostars because of the well-known problems with angular 
momentum dispersal (e.g., Bodenheimer 1995 and references therein). As we 
discussed above, our observational results alone are broadly consistent 
with this picture, with a total companion frequency of at least 
$\sim$65-80\%. In this section we will attempt to establish the frequency 
of single stars by estimating the frequency in the rest of parameter 
space, and ask if this is consistent with simple stochastic fragmentation 
models that neglect angular momentum considerations or feedback processes.

From our sample of 128 distinct, gravitationally bound systems (both 
binaries and singles), 11 systems are single to our knowledge but have no 
high-resolution observations, so we do not include them in the analysis of 
this section. Of the remaining 117 systems, 48 have no companions between 
3 and 5000\,AU, with detection limits reaching the substellar regime 
($q\la$0.1 or $M\la$100 $M_{Jup}$) in all cases and $\sim$10--20 $M_{Jup}$ 
for the majority of targets. This places an upper limit of 41\% for the 
single-star fraction of solar type stars in Taurus. To arrive at a true 
single-star fraction over all separations, we must estimate the number of 
close and wide binaries that we have missed.

The DM91 distribution converted to apparent separation in AU (at a total 
mass of 1\,M$_\odot$) has a mean in log of 1.39 and a standard deviation 
of 1.53. According to this distribution, 27\% of companions have 
separations smaller than 3\,AU. Conservatively using the DM91 
normalization of a 62\% companion fraction for $q>0.1$ systems leaves us 
with 8 of the 48 apparently single stars with close companions. An 
extrapolation of our separation distribution with $\mu(\log(\rho))=2.6$, 
$\sigma(\log(\rho))=1.5$ and $F=1.05$ would give 14 apparently single 
stars with close companions. Two of these are the known spectroscopic 
binaries DQ~Tau and V826~Tau.

For wide companions, we need to estimate the number of apparent binaries 
that are just chance alignments between non-gravitationally bound Taurus 
members. Over the separation range of 30-120 arseconds, the characteristic 
surface density of unbound neighbours is 40 deg$^{-2}$ (Kraus \& 
Hillenbrand 2008), meaning that 6$\pm$2 wide companions are actually just 
chance alignments. There are 10 apparent wide companions amongst our 48 
apparently single stars, so we assume that 4$\pm$2 of these are genuine 
physical companions. However, there could be additional undiscovered 
companions; several of these candidate ultrawide companions were 
discovered only within the past 1--2 years (Scelsi et al. 2008; Luhman et 
al. 2009).

Once close and wide systems are taken into account, we expect $\sim$30-38 
out of the 117 well-observed systems to be single, resulting in a single 
star fraction of 25-32\%. Despite the higher multiplicity of Taurus stars 
compared to the field, even the lower end of this range shows that it is 
not uncommon for stars to form alone in a low-density environment such as 
Taurus.

Our Bayesian analysis determined a model for the probability density of 
companions as a function of separation and mass-ratio. However, this does 
not directly determine the predicted fraction of single stars, or indeed 
the fraction of multiple systems of each order. In order to discuss these, 
we need to include prior constraints on some aspects of the semimajor axis 
and companion frequency probability distribution, because our data alone 
do not constrain the smallest and largest separations. We chose to use the 
projected separation distribution from Raghavan et al. (2010), with 
$\mu(\log(\rho))=1.6$ and $\sigma(\log(\rho))=1.5$. We have converted 
their period distribution to a projected separation distribution assuming 
a total mass of 1.5\,$M_{\odot}$ and assuming that projected separation is 
on average 0.8 times the semimajor axis (e.g. FM92). This distribution is 
very similar to that from DM91, and our discussion below is only weakly 
dependent on this prior assumption.

Using this separation distribution, our data give a companion fraction 
$F=1.15\pm0.15$ (Section 5.3; Figure 6). The simplest possible way to go 
from this probability density function to multiplicity is to assume that 
the likelihood of finding additional companions around a primary star at 
any separation is independent of previously found companions, and to 
neglect fragmentation of fragments. This would produce a multiplicity 
distribution which is a Poisson distribution with mean $F$. Such a simple 
assumption already gives a single star fraction of 32$\pm$5\% and a 
sextuple or higher fraction of 0.5$^{+0.3}_{-0.2}$\%, consistent with our 
single star fraction and the one sextuple system LkHa 332-G1 ABLkHa 
332-G2AB/V955 Tau AB.

A slightly more sophisticated argument must involve the possibility of 
hierarchical fragmentation. We model this in a Monte-Carlo method by 
treating our model probability density as a large-scale fragmentation 
probability. Each fragment can then re-fragment with the same probability 
density function, but only at a scale at least 3 times smaller than the 
previous fragmentation (to reflect the absence of stable orbits for 
nonhierarchical triples). The secondary has fragmentation suppressed with 
a probability of 25\% in order to represent the possibility of a fragment 
falling below our mass limit for primaries in our sample ($M_{frag}<0.25$ 
$M_{\sun}$; Section 2), and fragmentation ceases if two fragments have a 
semi-major axis in the 1-4\,arcsec range, to mimic our loss in sensitivity 
over this range (Section 2). This model results in the same single star 
fraction as the Poisson model (32$\pm$5\%), but a substantially higher 
sextuple or higher fraction of 4.5$^{+2.1}_{-1.7}$\%. Restricting the 
separation distribution to our observed 3--5000 AU range gives a single 
star fraction of 37$\pm$5\% and a sextuple (or higher) fraction of 
2.3$^{+1.2}_{-0.7}$\%.

This is consistent with the 48 apparently single stars and one quintuple 
or higher system in our sample of 117 stars. A more sophisticated 
treatment would need to take into account the mass-dependence of 
fragmentation. With this extensive data set, we find that there is nothing 
unusual about single stars, as there would be if angular momentum 
evolution required the formation of binary companions as part of 
protostellar collapse. We also find the existence of high-order multiple 
systems in Taurus to be a natural product of fragmentation in a 
dynamically pristine environment.

\section{Implications for (Multiple) Star Formation}

Binary formation is expected to occur via two complementary pathways; wide 
($\ga$100 AU) binary companions should form by fragmentation of the 
protostellar core during its initial freefall collapse (e.g. Bodenheimer 
\& Burkert 2001), while close ($\la$100 AU) binary companion can form via 
gravitational instability and fragmentation in the protostellar accretion 
disk of the primary star (Toomre 1964; Boss 2001; Clarke 2009), most 
likely modified by subsequent migration through the disk. Both processes 
are ultimated rooted in the need for a collapsing protostar to dispel its 
angular momentum, but otherwise the detailed physics are quite distinct. 
As we describe below, only some of the predictions from theoretical models 
are verified by our observations, which suggests that these models remain 
incomplete.

A successful model for star formation should include these processes and 
successfully match the observed properties of the binary star population. 
The newest generation of theoretical models now match the slope and 
turnover of the IMF (e.g., Bate 2009a), but requiring simultaneous 
agreement with the (potentially mass-dependent) frequency, separation 
distribution, and mass ratio distribution for binary systems is a far more 
demanding criterion, and one that has yet to be achieved. Any discrepancy 
with respect to observations will provide guidance in developing the next 
generation of models, marking the phenomena that might be lacking (i.e., 
radiative feedback or magnetic fields; Bate 2009b; Offner et al. 2009; 
Price \& Bate 2009) or overrepresented (dynamical interactions; Kraus \& 
Hillenbrand 2008, 2009a).

\subsection{The Primordial Separation Distribution}

The overall separation distribution for young ($\la$5 Myr) solar-type 
($M\sim$0.7--2.5 $M_{\sun}$) stars in loose associations is significantly 
different from that observed in the field. Past surveys (e.g., DM91; 
Raghavan et al. 2010) have suggested that the field distribution is 
unimodal and log-normal, with a mean separation of 30 AU. In contrast, our 
results show that the separation distribution for solar-mass stars is 
approximately log-flat over at least 3.5 decades of separation (3-5000 
AU), and our study of young star clustering (Kraus \& Hillenbrand 2008) 
suggests that the log-flat binary separation distribution might extend to 
at least 15000 AU. The uniformity of the separation distribution for 
solar-mass binaries is quite surprising; as we described above and in the 
introduction, binary formation should occur through very different 
processes at very large and very small separations. The presence of a 
discontinuity in binary properties near the expected transition point 
($\sim$100 AU) would confirm this expectation. The lack of a discontinuity 
does not necessarily disprove the expectation, but it does argue that both 
modes yield similar results despite the very different evolutionary paths.

For wide ($>>$100 AU) binary systems that are expected to fragment during 
or just after freefall collapse, the initial semimajor axis of the binary 
system should depend primarily on the characteristic size of the core when 
fragmentation occurred and the location within the core where the critical 
overdensity was reached. Taurus hosts solar-type binary systems with 
separations of up to $\sim 10^4$ AU, which is similar to the 
characteristic size of prestellar cores in regions like the Pipe Nebula 
that could resemble the Taurus progenitor (Lada et al. 2008). This 
similarity suggests that fragmentation can occur very early, before the 
outer envelope has undergone significant freefall collapse toward the 
central star. Observations of starless prestellar cores seem to indicate 
characteristic sizes of $\sim$10$^4$ AU (Menshchikov et al. 2010). Similar 
observations of more evolved Class 0 protostars seem to indicate that they 
have very large envelopes ($\ga$10$^3$ AU; Looney et al. 2000) and 
condense from the inside out, with the outer envelope remaining 
unperturbed (aside from some rotational flattening) while the inner 
envelope undergoes obvious infall (e.g., Chiang et al. 2010). Observations 
of Class I protostars in Taurus indicate that the outer envelope radius 
remains large (Furlan et al. 2008), but as we discuss in the next 
subsection, most companions must fragment before the primary accretes most 
of the envelope mass. We therefore suggest that binary fragmentation must 
occur no later than the Class 0 stage, especially since many of the Class 
0 systems observed by Looney et al. (2000) appear to have already 
fragmented into multiple widely-separated components by their present age.

Timing constraints aside, it is unclear how Jeans-critical fragments could 
initially form in the outer regions of protostellar cores. Observations 
show that prestellar cores can be approximated by pressure-confined, 
thermally-supported isothermal sphere (i.e., Bonnor 1956; Ward-Thompson et 
al. 1994) that have a density profile with approximately constant density 
for the inner $\sim$10$^3$ AU (where magnetic or turbulent support might 
dominate) and $\rho \propto r^{-2}$ at larger separations. In these cores, 
a Jeans-critical fragment would represent a far higher fractional 
overdensity at large radii as compared to small radii. The hydrodynamic 
models that are able to produce wide pairs (e.g., Delgado-Donate et al. 
2004) usually start with a uniform-density medium and allow the structure 
to emerge from free-fall and turbulent motions, but this is not consistent 
with the presence of quasistable pressure-confined cores even in pre-star 
forming environments like the Pipe Nebula (Lada et al. 2008). One solution 
might be for wide binary companions to fragment out of substructures that 
trace the larger structure of the star-forming region. Stars seem to form 
along large-scale filaments in their progenitor giant molecular cloud 
(Goldsmith et al. 2008; Menshchikov et al. 2010), so if protostellar cores 
remain elongated as they collapse (as might be suggested by observations; 
Tobin et al. 2010), then wide companions could form more easily along the 
filament direction. An observational test of this hypothesis would be to 
observe wide binaries among Class 0 stars that are $<<$1 orbital period 
old, and thus determine whether their PA is aligned with the local 
filamentary structure of the star-forming region. Many regions have been 
surveyed in the MIR with Spitzer (e.g. Rebull et al. 2010) and will be 
observed in the submm/mm with SCUBA-2 and ALMA, so this test could be 
feasible in the near future. Surveys of Class I binary systems (e.g. 
Connelley et al. 2008) might provide such a test, but most of the regions 
included in their sample have not been studied to characterize their 
larger-scale structure to the same extent as for Taurus (e.g. Goldsmith et 
al. 2008 for the gas or Kraus \& Hillenbrand 2008 for the stars).

It is also noteworthy that there are very few low-mass binary systems with 
very wide separations, but instead the separation distribution for 
lower-mass stars ($M_{prim}=$0.25--0.7 $M_{\sun}$) appears to be truncated 
for separations of $\ga$200 AU. This limit is similar to the limit seen in 
the field, where systems have been observed to follow a relation between 
the system mass $M$ and the maximum possible binary separation $a_{max}$. 
The functional form of this trend is $a_{max} \propto M^2$ for masses 
$<$0.4 $M_{\sun}$ (Burgasser et al. 2003), so the envelope corresponds to 
a constant binding energy at all masses. Previous studies have interpreted 
this binding energy cutoff as a signature of dynamical interactions, such 
that loosely bound systems are disrupted by interactions within the natal 
cluster. However, the interaction timescale in Taurus is much longer than 
its age, so external truncation by other stars does not seem to be a 
likely explanation. We therefore suggest that perhaps this correspondance 
with binding energy is coincidental across this regime, a point we made 
with respect to very wide binary systems in Kraus \& Hillenbrand (2009b), 
and that instead low-mass cores are simply unlikely to fragment during 
free-fall collapse. 

One alternate explanation which must be considered is that the outer 
separation limit is indeed dynamical in origin, but is set by the 
processes internal to the protostellar core and not by external stellar 
interactions. The most popular formulation of this concept is the ``embryo 
ejection'' model of star formation (Reipurth \& Clarke 2001), which 
postulates that protostellar cores produce many low-mass protostars or 
proto-brown dwarfs, but most are ejected from the protostellar envelope 
(and hence cut off from the reservoir of material to accrete) shortly 
after their fragmentation. This proposed mechanism originally seemed to be 
a natural complement to early gravoturbulent star formation simulations, 
which tended to fragment into dynamically active systems with many 
low-mass components. However, there have been ongoing debates regarding 
its observational predictions, particularly in the potential impact (or 
lack thereof) on the velocity and spatial distributions of star-forming 
regions (e.g. Luhman 2006; Kraus \& Hillenbrand 2008) and the frequency 
and properties of disks (e.g. White \& Basri 2003; Luhman 2004; Scholz et 
al. 2006). Recent updates to theoretical simulations may have solved these 
controversies; new model runs that incorporate magnetic fields and 
radiative feedback seem to produce systems with a few larger stars (Bate 
2009; Offner et al. 2009), rather than the many brown dwarfs seen in early 
models. We therefore suggest that even though some high-N systems appear 
to form (such as V773 Tau and V955 Tau), the many-body outcome of star 
formation might not be a representative case for collapse and 
fragmentation of a typical protostellar core.

Given the prevalence of binary companions at small separations, then disk 
fragmentation seems to remain a viable pathway for producing binary 
companions, so we instead suggest that the outer separation limit might 
indicate the maximum separation at which disk fragmentation can occur. For 
close ($<$100 AU) binary systems that are expected to form via disk 
fragmentation, the semimajor axis should depend on the radius at which 
fragmentation occurs and any subsequent migration of the binary companion. 
Disks are typically modeled using the formalism of $\alpha$-disk theory 
(Shakura \ Sunyaev 1973), which characterizes the viscosity as 
proportional to the local sound speed, vertical scale height, and a 
constant coefficient $\alpha$; the disk self-gravity is often modeled as a 
pseudo viscosity as well (e.g., Clarke 2009). The structure of early-stage 
protostellar disks is assumed to follow this model, but there are few 
observations that measure these disks' properties, so detailed predictions 
regarding the radius of initial fragmentation are not feasible yet. 
However, the most recent models predict that fragmentation should be most 
common in the outer portion of the disk ($\sim$50-100 AU), where cooling 
is more efficient and orbital shear is less important (Matzner \& Levin 
2005; Stamatellos et al. 2007).

We found many companions at separations down to $\sim$3 AU, so if the 
binary companions did not form in situ, then it seems likely that they 
formed at larger separations and migrated inward.  Any companion that 
forms via disk instability should be large enough to open a gap 
immediately, so subsequent migration should proceed via the Type II 
mechanism (Lin \& Papaloizou 1985) and carry the companion inward; this 
tendency is born out by simulations (Bate et al. 2002; Clarke et al. 
2009). The migration timescale depends on the primary and companion masses 
(being much longer for similar-mass companions) so it will depend on the 
accretion history of the system. For example, if the mass ratio is 1:100, 
then the migration timescale in a disk which is massive (10 times the 
Minimum-Mass Solar Nebula) and has a viscosity parameter of 
$\alpha$$=$10$^{-3}$ will be 10$^5$ yr at 1 AU and $5 \times 10^5$ yr at 
25 AU (Ida \& Lin 2004). Increasing the mass ratio by a factor of 10 will 
lengthen the migration timescale by a factor of 10, effectively freezing 
the companion at the location where significant accretion occurred. This 
suggests that the accretion history sets the final location of a 
companion, with the migration distance depending on the length of time 
before significant accretion occurs.

As we noted above, there are few low-mass binary systems 
($M_{prim}=$0.25--0.7 $M_{\sun}$) with separations of $\ga$200 AU, but 
binary systems are very common at smaller separations. This discrepancy 
seems to indicate that disk fragmentation could be the preferred mechanism 
for low-mass binary formation, with little contribution from early 
fragmentation during free-fall collapse. Since migration typically moves 
companions inward, the mass-dependent maximum separation for binary 
systems (Reid et al. 2001; Burgasser et al. 2003; Kraus \& Hillenbrand 
2009a) could then be interpreted as a denoting the separation regime at 
which disk fragmentation occurs: $\sim$100 AU for early M stars, 
$\sim$20-30 AU for late-M stars, and $\sim$5--10 AU for brown dwarfs.

\subsection{The Mass Ratio Distribution}

Our observed mass ratio distributions also do not agree with those of 
DM91, with a linear-flat primordial distribution in Taurus that features 
more similar-mass companions and fewer low-mass companions. There have 
been no processes suggested that would cause the field mass ratio 
distribution to differ from the young (i.e. Class II) mass ratio 
distribution, so given that DM91 relied on significant completeness 
corrections for the lowest-mass companions, we believe that the 
linear-flat distribution probably represents the true distribution in the 
field as well. As with the the separation distribution, we did not expect 
the observed similarity of the mass ratio distributions for wide 
companions ($\ga$100 AU) and very close companions ($\la$100 AU), so it is 
surprising that the different modes of binary formation yield similar 
results.

For wider binary pairs, fragmentation should cause the cloud to split into 
two separate core/envelope systems that then evolve independently, with 
some overlap of the envelopes for separations out to $\sim$10$^3$ AU. In 
this case, the division of mass between the two components should be 
driven by their relative locations within the progenitor core and the 
specific angular momentum of the remaining envelope. In the case of pure 
freefall collapse, material should tend to accrete onto the nearer 
fragment, which would tend to yield a flat mass ratio distribution if both 
positions are drawn at random. However, this should also yield a 
correlation such that wider binary systems have lower mass ratios, since 
fragments that initially form near the edge of the cloud will have access 
to less material to accrete. Conversely, if the material in the progenitor 
core has high specific angular momentum, then mass should preferentially 
accrete onto the star with the shallower potential well. This should drive 
the mass of the lower-mass secondary toward the mass of the primary, 
resulting in a mass ratio distribution that is peaked at unity.

Both of these results can be seen in hydrodynamic simulations (Bate et al. 
2000; Delgado-Donate et al. 2004; Bate 2009). However, we find no 
preference for similar-mass companions in this mass range, and as we 
showed for wide binaries in Kraus \& Hillenbrand (2009a), even the widest 
systems have a flat mass ratio distribution. More detailed observations of 
very young protostars during the epoch of fragmentation (e.g., Duch\^ene 
et al 2007) should cast more light on this process.

For close ($\la$100 AU) binary pairs, the secondary fragments out of a 
circumprimary accretion disk and the two stars share a common envelope. 
Dynamical constraints on the total mass of a self-gravitating disk ensure 
that the initial mass of the secondary will be less than the primary mass, 
but the masses should be significantly altered by subsequent accretion out 
of the envelope, and will ultimately be set by the total mass remaining to 
be accreted, the specific angular momentum of the envelope material, and 
the orbital radius of the companion. As we described above, the initial 
orbital radius of the companion should be large ($\sim$50--100 AU), but 
the radius will be modified (most likely inward) by migration and then 
frozen by additional accretion (e.g., Lin \& Papaloizou 1985; Bate et al. 
2002; Ida \& Lin 2004). The relative accretion onto the primary or 
secondary will then be set by the orbital radius at which envelope 
material accretes onto the disk (Clarke 2009), which will be determined by 
the specific angular momentum and is observationally measured to also have 
a characteristic radius of $\sim$50-100 AU (Watson et al. 2008). Any 
material which falls outside the companion should accrete onto it, while 
material that falls inside the companion should accrete onto the 
primary. Some simulations suggest the primary could see significant 
accretion even if the material has very high specific angular momentum 
(Ochi et al. 2005; Hanawa et al. 2010), though observations of systems 
with circumbinary disks tend to find that the secondary accretes more mass 
(e.g., Jensen et al. 2007).

Since observations indicate that envelope material accretes onto the disk 
at large radii, it seems likely that most of this material will accrete 
onto the secondary. The flat mass ratio distribution therefore indicates 
that at the time of companion fragmentation, the primary mass typically 
has approached its final mass and must constitute $\ga 1/2$ of the entire 
core mass; if the envelope is accreted at a time-averaged constant rate 
until it is depleted, then fragmentation must occur with constant 
probability at any point after this limit, such that companions have a 
flat distribution of masses up to the primary mass. If the companion 
fragmented earlier in the envelope accretion stage, then competitive 
accretion would drive the mass ratio to unity, after which subsequent 
accretion would occur equally onto either component, yielding a 
significant population of ``twins''. Conversely, if the companions tended 
to fragment significantly later, then there would be insufficient material 
for a significant fraction of all companions to grow to similar masses as 
their primary stars. Clarke (2009) has suggested that the delay in 
companion fragmentation could result from an initially compact 
configuration for typical protostellar disks. Viscous evolution will 
transport angular momentum outward in a disk, so its outer radius might 
not spread outward into the regime where fragmentation can occur 
($\sim$50--100 AU) until after a significant amount of material has been 
accreted. However, this model relies on the envelope containing low 
specific angular momentum, such that most accretion occurs at small 
separations. The small number of detailed studies that measure envelope 
accretion (e.g. Watson et al. 2007) suggest that it actually occurs at the 
same characteristic radius as fragmentation.

\subsection{The Frequency of Protostellar Fragmentation}

The high companion frequency for Taurus has led to a persistent meme that 
in sparse environments, nearly all stars are born with binary companions. 
Newly-formed stars must disperse a tremendous amount of angular momentum 
in condensing through $\sim$6 orders of magnitude in radius, so binary 
formation would offer a convenient sink for much of this excess angular 
momentum. High multiplicity among solar-type stars would also match the 
predictions of many gravoturbulent star formation models, which tend to 
form small-$N$ clusters that subsequently evolve into a high-mass multiple 
system and many low-mass single stars and brown dwarfs. However, the 
assertion of near-universal primordial multiplicity has not been tested 
across the wide separation range studied in our survey.

As we discuss in Section 5.4, it appears that $\sim$1/4--1/3 of all 
star-forming cores that can form at least one $>$0.25 $M_{\sun}$ star will 
yield only that one star. Binary systems therefore can not represent the 
only solution for overcoming the rotational support of angular momentum, 
though the single stars might represent the low-momentum tail of a natural 
distribution of total core angular momentum values. Our result shows that 
other processes like disk-locking (K\"onigl 1991; K\"onigl \& Pudritz 
2000; Rebull et al. 2006) and winds (Shu et al. 2000; Matt \& Pudritz 
2006) can be sufficient for dissipating a protostar's angular momentum. 
Our result also suggests that the dynamically active mode of 
gravoturbulent star formation has limited relevance to regions like 
Taurus; it is unlikely for solar-type stars to be rendered single in the 
decay of small-$N$ clusters (e.g., Goodwin et al. 2005), plus the low 
surface density and low velocity dispersion of Taurus members ($\Sigma \la 
5$ stars/pc$^2$ and $v\sim$200 m/s; Kraus \& Hillenbrand 2008) indicates 
that often there are no small-$N$ clusters from which these stars could 
have been ejected.

Rather than universal multiplicity, the high companion frequency in Taurus 
is reflected by the highly multiple nature of some systems.  In our 
subsample of 117 gravitationally bound systems with high angular 
resolution observations, there are 48 single stars, 50 binaries, 12 
triples, 4 quadruples and 1 sextuple when considering companions with 
separations of 3--5000 AU. The true multiplicities are higher than this: 
the J1-4872 system is a known quadruple, and the V807 Tau/GH Tau system is 
a known quintuple. However, one companion in each system was not included 
in our sample because we would not have been able to detect them with our 
survey, because of the AO "hole'' for binary separations between 1 and 
4\,arcsec. As we showed in Section 5.4, our Monte Carlo companion 
distribution code successfully replicates the distribution of high-order 
multiples by assuming that once fragmentation has occurred, each component 
can fragment on smaller scales at the same rate as a similar-mass core 
that had not fragmented. For comparison, our code predicts 5.9$\pm$2.4 
quintiple or higher order systems, once all completeness corrections are 
taken into account. This number would be even higher if fragmentation of 
systems with primary masses $<$0.25\,M$_\odot$ would be included. This is 
clearly higher than the number of high order multiples expected in a field 
population, and suggests that in typical clustered star formation 
environments, high-order multiples do not survive because wide pairs tend 
to be broken apart (e.g., Reipurth et al. 2007; Kraus \& Hillenbrand 
2009a).

\section{Summary}

We have conducted a high-resolution imaging study of the multiple star 
population in the Taurus-Auriga star-forming region. Our results have 
significant implications for the primordial outcome of multiple star 
formation. To summarize:

\begin{enumerate}

\item We have identified 16 additional binary companions to primary stars 
with masses of 0.25-2.5 $M_{\sun}$, raising the total number of companions 
at separations 0.015--30\arcsec\, to 90. Combined with our previous survey 
of wide binary systems, we have now compiled a comprehensive census 
spanning separations of 3--5000 AU.

\item We have found that $\sim$2/3--3/4 of all Taurus targets are multiple 
systems of two or more stars, while the other $\sim$1/4--1/3 appear to 
have formed as single stars. The distribution of high-order multiples is 
consistent with fragmentation occurring independently on all scales; once 
a collapsing protostellar core has fragmented into two components, either 
component can further fragment with the same probability as a single star 
of the same mass.

\item For solar-type stars (0.7--2.5 $M_{\sun}$), the separation 
distribution is very nearly log-flat over separations of 3--5000 AU. In 
contrast, lower-mass stars (0.25--0.7 $M_{\sun}$) show a paucity of binary 
companions with separations of $\ga$200 AU. Across this full mass range, 
the companion mass function is well described as a linear-flat function; 
all companion mass ratios are equally probable, apparently including 
substellar companions.

\item Binary formation on large scales ($\ga$100-200 AU) probably 
occurs via fragmentation during initial free-fall collapse, so the 
existence of wide companions (extending to $\sim$10$^4$ AU) indicates that 
fragmentation can occur very early. We suggest that these companions might 
find their origin in the traces of larger-scale structure within the 
cloud, as a spherically symmetric isothermal sphere should not easily 
fragment on these scales, and if it did, it would preferentially form 
lower-mass companions. The paucity of wider companions to low-mass 
primaries might indicate that low-mass protostellar cores do not fragment 
during freefall collapse.

\item Binary formation on smaller scales ($\la$100-200 AU) probably occurs 
via fragmentation of the protostellar accretion disk that forms after 
freefall collapse has ended. Fragmentation should occur in the outer disk 
($\ga$50 AU), so the log-flat separation distribution must indicate how 
far these companions migrate inward before they grow too massive to 
migrate. The flat mass ratio distribution seems to indicate that 
fragmentation occurs during the last half of envelope accretion. If 
fragmentation occurred during while most mass was still in the envelope, 
then competitive accretion would drive system mass ratios preferentially 
to unity. Conversely, if most fragmentation occurred late, then the 
envelope would lack sufficient mass to grow $\sim$1/2 of all companions to 
within $\ga$1/2 of the primary mass.
 
\end{enumerate}

\acknowledgements

We thank Peter Tuthill and Jamie Lloyd for campaigning to have aperture 
masks installed in PHARO and NIRC2. We also thank Russel White for sharing 
the results of his multiplicity survey in Taurus prior to publication. 
Finally, we thank the referee for a detailed critique of this paper. ALK 
has been supported by a SIM Science Study and by NASA through Hubble 
Fellowship grant 51257.01 awarded by STSCI, which is operated by AURA, 
Inc., for NASA, under contract NAS 5-26555. MI is supported by an 
Australian Postdoctoral fellowship from the Australian Research Council. 
This work makes use of data products from 2MASS, which is a joint project 
of the University of Massachusetts and IPAC/Caltech, funded by NASA and 
the NSF. Our research has also made use of the USNOFS Image and Catalogue 
Archive operated by the United States Naval Observatory, Flagstaff Station 
(http://www.nofs.navy.mil/data/fchpix/).

We recognize and acknowledge the very significant cultural role and 
reverence that the summit of Mauna Kea has always had within the 
indigenous Hawaiian community. We are most fortunate to have the 
opportunity to conduct observations from this mountain.

\clearpage
\LongTables

\begin{deluxetable*}{lllcrrrl}
\tabletypesize{\scriptsize}
\tablewidth{0pt}
\tablecaption{Stars in Taurus-Auriga with SpT$\le$M4}
\tablehead{\colhead{Name} & \colhead{RA} & \colhead{DEC} & \colhead{SpT} & 
\colhead{$R$} & \colhead{$K$} & 
\colhead{Group\tablenotemark{a}} & \colhead{Refs}
\\
\colhead{} & \multicolumn{2}{c}{(J2000)} & \colhead{} &
\colhead{(mag)} & \colhead{(mag)}
}
\startdata
\multicolumn{8}{l}{\bf Observed Members}\\
2M04080782&4 08 07.82&28 07 28.0&M3.75&15.0&11.39&P&1\\
LkCa 1&4 13 14.14&28 19 10.8&M4&12.6&8.63&9, H&2\\
Anon 1&4 13 27.23&28 16 24.8&M0&12.1&7.46&F&2\\
FM Tau&4 14 13.58&28 12 49.2&M0&13.3&8.76&12, L&2\\
CW Tau&4 14 17.00&28 10 57.8&K3&12.3&7.13&8&2\\
FP Tau&4 14 47.31&26 46 26.4&M4&12.7&8.87&11&2\\
CX Tau&4 14 47.86&26 48 11.0&M2.5&12.6&8.81&10, K&2\\
\enddata
\tablecomments{Positions and K magnitudes are from 2MASS, R 
magnitudes are from NOMAD (Zacharias et al. 2004). Magnitudes 
are uncertain by $\sigma_R \sim$0.25 mag and $\sigma_K 
\sim$0.03 mag. Spectral types and membership are determined 
from the references listed. References: 
1) Luhman (2006); 
2) Kenyon \& Hartmann (1995); 
3) Strom \& Strom (1994);
4) G. Herczeg, priv. comm.;
5) White \& Hillenbrand (2004);
6) Duch\^ene et al.(1999);
7) Hartigan et al. (1994); 
8) Kraus \& Hillenbrand (2008);
9) White \& Hillenbrand (2005);
10) Brice\~no et al. (1999);
11) Brice\~no et al. (1998);
12) Hartigan \& Kenyon (2003);
13) White et al. (1999);
14) White \& Basri (2003);
15) Schaefer et al. (2009);
16) Luhman \& Rieke (1998);
17) Luhman (2000);
18) Kenyon et al. (1998);
19) Brice\~no et al. (2002);
20) Luhman et al. (2003);
21) Mart\'in et al. (2000);
22) Slesnick et al. (2006);
23) Luhman (2004);
24) NOMAD (Zacharias et al. 2004);
25) White \& Ghez (2001);
26) Stapelfeldt et al. (2003).
}
 \tablenotetext{a}{Our survey targets were observed in groups, such that 
single targets could be used as calibrators for newly-discovered binaries. 
We also list the composition of these groups in Table 2.}
 \tablenotetext{b}{Short-period pectroscopic binary, as characterized by 
Mathieu et al. (2007) for DQ Tau AB, Prato et al. (2002) for UZ Tau Aab, 
White \& Hillenbrand (2005) for StHa 34 Aab, and Massarotti et al. (2005) 
for V826 Tau AB. For our analysis, the component masses of spectroscopic 
binaries will be combined and treated as a single higher mass.}
 \end{deluxetable*}

\clearpage

\begin{deluxetable*}{rcccl}
\tabletypesize{\scriptsize}
\tablewidth{0pt}
\tablecaption{Observed Groups}
\tablehead{\colhead{Group} & \colhead{Telescope} &
\colhead{Filter} & \colhead{Epoch} & \colhead{Stars}
\\
\colhead{} & \colhead{} & \colhead{} & 
\colhead{(MJD)} & \colhead{}
}
\startdata
\multicolumn{5}{l}{\bf Fall 2007}\\
1&Keck&Lp&54428.3&RY Tau, SU Aur, HP Tau-G2, AB Aur, HD 283572\\
2&Keck&Kp&54427.3&GM Aur, UX Tau A, LkCa 14, LkCa 19\\
3&Pal&Ks&54432.3&HBC 388, DR Tau, HBC 427\\
4&Pal&Ks&54434.3&V827 Tau, L1551-51, V826 Tau, DN Tau, DS Tau\\
5&Pal&Ks&54434.3&HBC 352, HBC 353, BP Tau, DG Tau\\
6&Keck&Kp&54427.3&LkCa4, DE Tau, LkCa 15, V830 Tau\\
7&Keck&Kp&54427.5&GK Tau, GI Tau, DL Tau, Hubble 4\\
\enddata
\end{deluxetable*}

\clearpage

\begin{deluxetable*}{lclcccc}
\tabletypesize{\scriptsize}
\tablewidth{0pt}
\tablecaption{Resolved Binary Companions}
\tablehead{\colhead{Primary} & \colhead{Telescope} 
& \colhead{Filter} & \colhead{Epoch} & 
\colhead{Separation} & \colhead{P.A.} & \colhead{$\Delta$$m$}
\\
\colhead{} & \colhead{} & \colhead{} & \colhead{(MJD)} & 
\colhead{(mas)} & \colhead{(deg)} & \colhead{(mag)}
}
\startdata
2M04080782&Keck&Kp&54822.5&44.0$\pm$0.9&133.1$\pm$0.8&2.67$\pm$0.04\\
Anon 1&Keck&Kp&54821.3&14.9$\pm$0.5&144$\pm$3&0.46$\pm$0.09\\
LkCa 5&Keck&Kp&54427.4&48$\pm$4&61$\pm$3&2.89$\pm$0.16\\
Hubble 4&Keck&Kp&54427.5&28.4$\pm$0.1&106.07$\pm$0.13&0.40$\pm$0.01\\
LkCa 21&Keck&Kp&54428.5&44.4$\pm$0.1&113.6$\pm$0.2&0.35$\pm$0.01\\
J1-507&Keck&Kp&54823.3&79.4$\pm$0.3&19.22$\pm$0.13&0.03$\pm$0.01\\
V827 Tau&Palomar&H&54434.3&92.9$\pm$0.3&14$\pm$0.2&0.58$\pm$0.01\\
FF Tau&Keck&Kp&54427.6&36.3$\pm$0.4&356.4$\pm$0.5&1.03$\pm$0.02\\
HP Tau-G3&Keck&Kp&54427.6&30.5$\pm$1.5&94.2$\pm$1.2&1.44$\pm$0.14\\
HV Tau&Keck&Kp&54821.5&36.0$\pm$0.2&326.6$\pm$0.3&0.63$\pm$0.01\\
CoKu Tau/4&Keck&Kp&54427.6&54.1$\pm$0.3&306.7$\pm$0.4&0.23$\pm$0.01\\
2M04414565 A&Keck&Kp&54822.3&224.0$\pm$0.4&87.10$\pm$0.11&2.96$\pm$0.03\\
DP Tau&Keck&Kp&54428.5&106.7$\pm$0.8&293.3$\pm$0.3&0.36$\pm$0.02\\
HBC 427&Keck&Kp&54823.5&32.3$\pm$0.1&180.24$\pm$0.13&0.87$\pm$0.01\\
CIDA-10&Keck&Kp&54822.3&83.0$\pm$0.2&60.47$\pm$0.22&0.01$\pm$0.01\\
CIDA-11&Keck&Kp&54822.5&97.2$\pm$0.2&277.66$\pm$0.08&0.31$\pm$0.01\\
\enddata
\end{deluxetable*}

\clearpage

\begin{deluxetable*}{lcccllllll}
\tabletypesize{\scriptsize}
\tablewidth{0pt}
\tablecaption{Companion Detection Limits}
\tablehead{\colhead{Primary} & \colhead{Telescope} & 
\colhead{Filter} & \colhead{Epoch} & 
\multicolumn{6}{c}{$\Delta$$m$ (mag) at $\rho=$ (mas)} 
\\
\colhead{} & \colhead{} & \colhead{} & \colhead{(MJD)} & 
\colhead{10-20\tablenotemark{a}} & \colhead{20-40} & \colhead{40-80} & \colhead{80-160} & 
\colhead{160-240} & \colhead{240-320} 
}
\startdata
HBC 352&Palomar&Ks&54434.3&0.00&0.00&2.69&3.97&4.09&4.06\\
HBC 353&Palomar&Ks&54434.3&0.00&0.00&2.05&3.58&3.70&3.69\\
2M04080782&Keck&Kp&54822.5&0.99&3.21&4.61&4.58&4.23&4.17\\
LkCa 1&Keck&Kp&54427.4&0.00&1.12&3.03&2.93&4.33&4.36\\
LkCa 1&Keck&Kp&54823.5&3.41&5.08&6.01&5.80&4.58&4.54\\
Anon 1&Keck&Kp&54821.3&1.95&3.82&5.13&5.07&5.69&5.74\\
FM Tau&Keck&Kp&54428.6&0.00&2.19&3.67&3.49&4.04&3.99\\
FM Tau&Keck&Kp&54823.3&1.77&3.77&5.13&5.04&4.28&4.24\\
\enddata
\tablenotetext{a}{Each column reports the detection limit (in terms of 
flux ratio $\Delta m$) for a bin with the given range of projected 
separations (in mas).}
\end{deluxetable*}

\clearpage

\begin{deluxetable*}{lllllccl}
\tabletypesize{\scriptsize}
\tablewidth{0pt}
\tablecaption{Derived Properties for Class II/III Binaries in Taurus}
\tablehead{\colhead{Primary} & \colhead{Secondary} & 
\colhead{$M_{prim}$\tablenotemark{a}} & 
\colhead{$M_{sec}$\tablenotemark{a}} & 
\colhead{$q$\tablenotemark{b}} & \colhead{Sep\tablenotemark{c}} 
& \colhead{$q$} & 
\colhead{Refs}
\\
\colhead{Name} & \colhead{Name} & \colhead{($M_{\odot}$)} & 
\colhead{($M_{\odot}$)} & \colhead{($M_{s}/M_{p}$)} & \colhead{(AU)} & 
\colhead{Type\tablenotemark{b}} 
}
\startdata
2M04080782 A&2M04080782 B&0.30&(0.04)&0.13&6.4&$\Delta$$K$&1\\
Anon 1 A&Anon 1 B&0.64&(0.43)&0.68&2.2&$\Delta$$K$&1\\
MHO 2 A&MHO 2 B&0.33&(0.11)&0.34&7.3&$\Delta$$K$&2\\
MHO 3 A&MHO 3 B&0.72&(0.36)&0.50&4.5&$\Delta$$K$&2\\
LkCa 3 A&LkCa 3 B&0.57&(0.46)&0.81&69&$\Delta$$K$&3\\
FO Tau A&FO Tau B&0.33&(0.33)&1.00&22&$\Delta$$K$&3\\
LkCa 5 A&LkCa 5 B&0.50&(0.05)&0.09&6.9&$\Delta$$K$&1\\
V410 Tau A&V410 Tau B&0.94&(0.19)&0.20&17.8&$\Delta$$K$&4\\
\enddata
\tablecomments{References: 
1) This work;
2) White et al. (2011);
3) White \& Ghez (2001);
4) Ghez et al. (1997);
5) Kraus \& Hillenbrand (2007a);
6) Correia et al. (2006);
7) Chen et al. (1990);
8) Woitas et al. (2001);
9) Ghez et al. (1993);
10) Duch\^ene et al. (1999);
11) Simon et al. (1995,1996);
12) Sartoretti et al. (1998);
13) Leinert et al. (1993);
14) Woitas \& Leinert (1998);
15) Ireland \& Kraus (2008);
16) Richichi et al. (1999).
}
 \tablenotetext{a}{The primary mass was determined from its corresponding 
spectral type, as described in Section 3.3. A majority of the secondary 
masses were determined from the primary mass and the calculated mass ratio 
(quantities in parentheses), but many were directly calculated from 
measured spectral type of the secondary component. The uncertainties are 
likely to be $\sim$20\% and are set by systematic uncertainties from the 
models and the highly uncertain ages of Taurus members.}
 \tablenotetext{b}{As we describe in Section 3.3, the mass ratio of each 
system was estimated from the mass-spectral type relation described in the 
text (SpT-M), a dynamical mass measurement (DynM), or from the flux ratio 
($\Delta I$, $\Delta H$, or $\Delta K$). For some hierarchical multiple 
systems, several techniques were required to estimate all component 
masses. The uncertainties are likely to be $\sigma_{q}\sim$0.1, but could 
be much larger if there are edge-on disks or unresolved components that 
have not been identified.}
 \tablenotetext{c}{As we discuss in Section 3.3, projected physical 
separations are determined from the projected angular separations and the 
mean distance to Taurus. Projected physical separations are uncertain by 
$\sim$15\% due to the finite spread of distances to Taurus members 
($\sim$20 pc).}
 \end{deluxetable*}

\clearpage

\begin{landscape}

\begin{deluxetable}{llllllllllllllllllllll}
\tabletypesize{\tiny}
\tablewidth{0pt}
\tablecaption{Compiled Detection Limits for Companions}
\tablehead{\colhead{Primary} & \colhead{Mass} & 
\multicolumn{18}{c}{$q_{lim}$ ($m_s/m_p$) at $\rho =$ (mas)} & 
\colhead{Refs}
\\
\colhead{} & \colhead{($M_{\odot}$)} & 
\colhead{13.8\tablenotemark{a}} & \colhead{17.3} & \colhead{21.8} & 
\colhead{27.5} & \colhead{34.6} & \colhead{43.5} & 
\colhead{54.8} & \colhead{69.0} & \colhead{86.8} & 
\colhead{138} & \colhead{218} & \colhead{346} & 
\colhead{690} & \colhead{1376} & \colhead{2746} & 
\colhead{5478} & \colhead{10930} & \colhead{27455}
}
\startdata
2M04080782 A&0.3&0.282&0.177&0.093&0.075&0.053&0.036&0.036&0.036&0.036&1.00&1.00&1.00&1.00&1.00&1.00&1.00&1.00&1.00&1\\
2M04080782 AB&0.34&1.00&1.00&1.00&1.00&1.00&1.00&1.00&1.00&1.00&0.031&0.036&0.125&0.125&0.125&0.102&0.102&0.102&0.102&1\\
LkCa 1&0.27&0.066&0.042&0.024&0.021&0.018&0.015&0.015&0.015&0.015&0.017&0.029&0.148&0.148&0.115&0.037&0.018&0.018&0.018&1\\
Anon 1 A&0.64&0.131&0.087&0.051&0.042&0.03&0.02&1.00&1.00&1.00&1.00&1.00&1.00&1.00&1.00&1.00&1.00&1.00&1.00&1,2\\
Anon 1 AB&1.07&1.00&1.00&1.00&1.00&1.00&1.00&0.012&0.012&0.012&0.013&0.009&0.067&0.009&0.009&0.009&0.006&0.006&0.006&1,2\\
FM Tau&0.64&0.157&0.099&0.053&0.043&0.03&0.02&0.02&0.02&0.021&0.022&0.037&0.111&0.111&0.082&0.038&0.016&0.016&0.016&1\\
CW Tau&0.94&0.266&0.149&0.058&0.051&0.041&0.034&0.034&0.034&0.034&0.036&0.038&0.047&0.047&0.047&0.034&0.006&0.006&0.006&1,3\\
MHO-2 A&0.33&1.00&1.00&1.00&1.00&1.00&0.175&0.143&0.133&0.121&0.104&1.00&1.00&1.00&1.00&1.00&1.00&1.00&1.00&4,5\\
\enddata
\tablenotetext{a}{Each column reports the detection limit (in terms of 
companion mass ratio) for a bin with the given projected separation (in 
mas).}
\tablecomments{All observation limits include our wide binary results from 
Kraus \& Hillenbrand (2007a, 2009a). Other references:
1) This Work;
2) Tanner et al. (2007);
3) Ghez et al. (1993);
4) Duch\^ene et al. (2007);
5) White et al. (2010);
6) Sartoretti et al. (1998);
7) Leinert et al. (1993);
8) Massarotti et al. (2005);
9) Simon et al. (1995);
10) Konopacky et al. (2007);
11) Correia et al. (2006);
12) Simon et al. (1999);
13) Richichi et al. (1999).
}
\end{deluxetable}

\clearpage
\end{landscape}

\end{document}